\documentclass[12pt]{article}
\pdfoutput=1
\usepackage{multirow}
\usepackage{calc}
\usepackage{amsmath,amssymb,amsthm,amscd}
\numberwithin{equation}{section}
\usepackage[bf]{caption}
\usepackage{longtable}
\usepackage{array}
\usepackage{enumerate}
\usepackage{float}
\usepackage{subfig}
\usepackage{multicol}
\usepackage{multirow}
\usepackage{epsfig}
\usepackage{graphicx}
\usepackage{pict2e}
\usepackage{color}
\usepackage{authblk}
\usepackage{cite}
\usepackage[all]{xy}
\usepackage[width=1.09\textwidth]{caption}
\usepackage{bbm}
\usepackage{hyperref}

\usepackage{tikz}
\usepackage{tikz-cd}

\usepackage[utf8]{inputenc}
\usepackage[T1]{fontenc}
\usepackage{xspace}
\usepackage{pifont}
\usepackage[OT2,T1]{fontenc}

\DeclareSymbolFont{cyrletters}{OT2}{wncyr}{m}{n}
\DeclareMathSymbol{\Sha}{\mathalpha}{cyrletters}{"58}

 \makeatletter
\newcommand\xleftrightarrow[2][]{%
  \ext@arrow 9999{\longleftrightarrowfill@}{#1}{#2}}
\newcommand\longleftrightarrowfill@{%
  \arrowfill@\leftarrow\relbar\rightarrow}
\makeatother

\newcommand{\beq}{\begin{equation}}
\newcommand{\eeq}{\end{equation}}
\newcommand{\bea}{\begin{eqnarray}}
\newcommand{\eea}{\end{eqnarray}}

\newcommand{\set}[1]{\left\{ #1\right\}}

\renewcommand{\AA}{\mathbb{A}}

\newcommand{\CC}{\mathbb{C}}

\newcommand{\FF}{\mathbb{F}}

\newcommand{\PP}{\mathbb{P}}
\newcommand{\QQ}{\mathbb{Q}}

\newcommand{\ZZ}{\mathbb{Z}}

\newcommand{\al}{\alpha}

\newcommand{\om}{\omega}

\newcommand{\si}{\sigma}

\newcommand{\mx}{\mathfrak{m}}

\newcommand{\Hh}{\mathcal{H}}

\newcommand{\Ll}{\mathcal{L}}

\newcommand{\Oo}{\mathcal{O}}

\usepackage{bbm}

\newcommand{\spec}{\mathrm{Spec}}

\newcommand{\inv}{^{-1}}

\newcommand{\ip}[1]{\left\langle #1 \right\rangle }

\newcommand{\ttm}[4]{\begin{pmatrix} #1 & #2 \\ #3 & #4 \end{pmatrix}}

\newtheorem{thm}{Theorem}[section]
\newtheorem{df}[thm]{Definition}

\newtheorem{lem}[thm]{Lemma}
\newtheorem{prp}[thm]{Proposition}
\newtheorem{cor}[thm]{Corollary}

\begin{document}

\baselineskip=15pt
\begin{titlepage} 
\begin{center}
\vspace*{ 2.0cm}
{\Large {\bf Modular Curves and Mordell-Weil Torsion in F-theory}}\\[12pt]
\vspace{-0.1cm}
\bigskip
\bigskip 
{
{{Nadir Hajouji}$^{\,\text{a}}$} and {{Paul-Konstantin~Oehlmann}$^{\,\text{b}}$}
\bigskip }\\[3pt]
\vspace{0.cm}
{\it
  ${}^{\text{a}}$ Departments of Mathematics,~University of California Santa Barbara,~Santa Barbara,~CA 93106,~USA\\
 ${}^{\text{b}}$ Physics Department,~Robeson Hall,~Virginia Tech,~Blacksburg,~VA 24061,~USA  
}
\\[2.0cm]
\end{center}

\begin{abstract}
\noindent
In this work we prove a bound for the torsion in Mordell-Weil groups of smooth elliptically fibered Calabi-Yau 3- and 4-folds. In particular, we show that the set which can occur on a smooth elliptic Calabi-Yau $n$-fold for ($n\geq 3$) is contained in the set of subgroups which appear on a rational elliptic surface, and is slightly larger for $n = 2$. The key idea in our proof is showing that any elliptic fibration with sufficiently large torsion has singularities in codimension 2 which do not admit a crepant resolution. 
We prove this by explicitly constructing and studying maps to a modular curve whose existence is predicted by a universal property.
We use the geometry of modular curves to explain the minimal singularities that appear on an elliptic fibration with prescribed torsion, and to determine the degree of the fundamental line bundle (hence the Kodaira dimension) of the universal elliptic surface which we show to be consistent with explicit Weierstrass models. The constraints from the modular curves are used to bound the fundamental group of any gauge group $G$ in a supergravity theory obtained from F-theory. We comment on the isolated 8-dimensional theories, obtained from extremal K3's, that are able to circumvent lower dimensional bounds. These theories neither have a heterotic dual, nor can they be compactified to lower dimensional minimal SUGRA theories. We also comment on the maximal, discrete gauged symmetries obtained from certain Calabi-Yau threefold quotients. 
\end{abstract}

\end{titlepage}
\clearpage
\setcounter{footnote}{0}
\setcounter{tocdepth}{2}
\tableofcontents
\clearpage

\section{Introduction}
\label{sec:intro}
Calabi-Yau manifolds play a central role in the description of lower dimensional field and supergravity (SUGRA) theories through their use as compactification backgrounds of String/M/F-theory. 
Properties of the effective field theories (EFT) such as matter and gauge theory content  descend directly from geometric properties of the compactification spaces. The set of all EFTs, that can be obtained via string compactifications is refereed to as the {\it string landscape}.  The string landscape is distinguished from the so called {\it swampland}  \cite{Vafa:2005ui},
the set of only seemingly consistent EFTs that do not have a consistent high energy completion together with gravity  (for a review see \cite{Palti:2019pca}) and therefore not a geometric description as a string compactification. 
 This makes a full classification and study of the string landscape 
an interesting program both for physics and mathematics. 
The goal of the program is to find the rules that allow one to distinguish the swampland from the landscape. 
In the past, these {\it quantum gravity conjectures} existed often as  folk  theorems making use of heuristic black hole arguments, which by themselves   might not be fully understood.\\
In recent years a lot of a attention has been given to finding and connecting the set of {\it quantum gravity conjectures} and developing tools to put them on more solid grounds, as for example happened for {\it weak gravity} \cite{ArkaniHamed:2006dz,Lee:2018urn} and {\it distance conjectures} \cite{Ooguri:2006in,Klaewer:2016kiy,Grimm:2018ohb}. These tools make use of the geometric description of physics via Calabi-Yau compactifications and their geometric properties. \\ 
The framework of F-theory \cite{Vafa:1996xn} extends the geometrization of physics,
enabling descriptions of the largest patch of the string vacua (see e.g. \cite{Taylor:2015xtz}) in a single framework to date. 
The key idea of F-theory is to focus on elliptically fibered Calabi-Yau manifolds $Y$ and to reformulate the monodromies of $[p,q]$ 7-branes that wrap suitable cycles in the base $B$ acting on the IIB axio-dilaton into a geometric action on the complex structure of the elliptic fiber. The gauge algebra living on the branes is encoded in the singularity types of the elliptic fiber and goes beyond the limits of perturbative type IIB.\\ 
Certain global aspects of the gauge group $G$ of the F-theory compactifications are encoded in the Mordell-Weil (MW) group $MW(Y/B)$ of the fibration $Y$.
 For fibrations with a varying axio-dilaton, the MW group  admits  a decomposition:
\begin{align}
MW(Y/B)=\mathbb{Z}^r \times T \, ,
\end{align}
where either $T \cong \ZZ_n$ or $T \cong \ZZ_n \times \ZZ_m$ for some integers $n,m, r$. We refer to $T$ is the torsion subgroup of the Mordell-Weil group. 
A priori\footnote{That is, without making assumptions about the canonical bundle of the total space.}, there are no constraints on the possibilities for $n, m, r$: 
any combination can, in principle, be obtained by pulling back the fibration
along a suitable map $B'\rightarrow B$.
Classical results on elliptic surfaces show that assumptions about the canonical bundle of the total space constrain the possibilities for the Mordell-Weil group when $\dim Y = 2$.  
In that case, a complete classification of Mordell-Weil groups and configurations of singular fibers is known for rational elliptic surfaces and for elliptically fibered K3s (see e.g. \cite{mirandaconfigs,perssonconfig,shimadak3} as well as \cite{Huybrechts:2016uxh} and references therein). 
A table with the torsion subgroups that occur on rational surfaces and K3 surfaces can be found in \cite{MP}.
There are no analogous classifications for smooth Calabi-Yau elliptic 3-folds and higher.   The present work is a step into that direction, by putting bounds on the torsion subgroup $T$. 

Within M-and F-theory, the rank $r$ of the MW group gives rise to $r$ U(1) gauge symmetry factors. This has been extensively studied in the context of F-theory \cite{Mayrhofer:2012zy,Morrison:2012ei,Morrison:2014era,Braun:2013nqa,Cvetic:2013nia,Cvetic:2018bni}.  
Abelian symmetries are in fact quite subtle. 
To constraint their maximal possible number $r$ with the helpf of the strong 6d SUGRA anomalies is only partially possible \cite{Park:2011wv} whereas the range of possible U(1) singlet charges cannot be constrained at all \cite{Taylor:2018khc}.  
 In this regard, the geometric description of Abelian symmetries via the free  part of the MW group allows for a more systematic and consistent exploration of matter charges \cite{Lawrie:2015hia,Raghuram:2017qut,Raghuram:2018hjn,Cianci:2018vwv,Collinucci:2019fnh} and maximal Abelian factors \cite{Lee:2019skh}. \\
On the other hand, 
the torsion part of the MW group
has been given much less attention in the recent literature. 
In F-theory compactifications, its effect can be understood as a refinement of the co-weight lattice of representations \cite{Aspinwall:1998xj,Mayrhofer:2014laa,Oehlmann:2016wsb,Esole:2017hlw} geometrically induced by the {\it torsion Shioda map}: The torsion Shioda map is in correspondence to a fractional linear combination of Cartan generators of the gauge algebra $\mathcal{G}$ that lives on stacks of seven branes. This allows to define a center charged for any matter representation $\mathbf{R}$ of $\mathcal{G}$ which must be integral valued to be compatible with the global gauge group $G$ . Although in F-theory only the massless matter spectrum is directly visible, one might expect this constraint to apply to the massive sector as well. Such situations can be viewed as the gauging of $\mathcal{G}$ by a discrete symmetry \cite{Aharony:2013hda} resulting in a non simply connected gauge group $G$ with fundamental group $\pi_1(G)=T$. A similar effect can happen, although of slightly different origin, when additional U(1) factors embed non-trivially inside the center of other non-Abelian group factors. In the following however we want to focus purely on the non-Abelian case without additional Abelian gauge factors.\\
  With a view on the swampland program, this begs the obvious question, which non-simply connected groups can be consistently realized in a quantum gravity, specifically in F-theory compactifications. In particular one might ask for bounds on the fundamental group $\pi_1 (G)$, due to the expectation that the string landscape must be finite\footnote{The set of elliptically fibered CY threefolds, and therefore landscape of 6D F-theory compactifications in fact has been proven to be ``bounded", in the sense that there all such threefolds fit into one of finitely many families, see \cite{Gross93}.}.   \\ 
    In \cite{Aspinwall:1998xj} Aspinwall and Morrison list various Weierstrass models featuring a wide variety of Mordell-Weil torsion subgroups, going up to $\ZZ_6$ for cyclic groups and $\ZZ_3 \times \ZZ_3$ in general. From the perspective of pure 6D SUGRA, it appears possible to set up models whose massless spectrum is consistent with (almost) any putative fundamental group $\pi_1(G)=\mathbb{Z}_n$ factor, where $n$ goes way beyond six. However, since the massless fields, are only a small sector of the full theory, one should be careful to draw conclusions about symmetries of the full theory, without having some good candidate entity that controls it. In F-theory, there exists such such an entity in terms of the torsion sections and their effect via the torsion Shioda map. In fact, the mere presence of torsion sections forces the existence of a minimal configuration of singularities whose structure often comes as a surprise, especially for higher order torsion. 
    \\\\
In this paper, we want to prove that the list of MW torsion models appearing in Aspinwall and Morrison \cite{Aspinwall:1998xj} contains every group that can be realized on a smooth, Calabi-Yau $n$-fold with $n >2$.
This geometric result allows us to put sharp {\it swampland} constraints on the fundamental group of gauge groups in SUGRA theories constructed from $[p,q]$-7 branes in F-theory. As a byproduct, we give a new perspective on elliptic fibrations with torsion sections by connecting them to the modular curve of certain congruence subgroups of $\mathrm{SL}(2,\mathbb{Z})$.\footnote{In a related F-theory context, fibrations with restricted monodromies have also been considered in \cite{Berglund:1998va,Bershadsky:1998vn} and recently \cite{Cota:2019cjx}.} These modular curves enforce a certain configuration of singularities on any elliptic fibration with a prescribed torsion group, which can then allow one to prove those torsion groups can't be realized on smooth Calabi-Yau's.\\
  In a different context, torsion sections in the Mordell-Weil group can be used as a building block to construct the covering geometry of a specific class of smooth, non-simply connected Calabi-Yau quotient torsors \cite{Donagi:1999ez}. Their associated F-theory physics admits discrete symmetries coupled to superconformal matter and gravity \cite{DelZotto:2014fia,Anderson:2018heq,Anderson:2019kmx} of the same order as the torsion factor of the covering geometry. Therefore, our bounds also translate to bounds on manifolds that can be obtained using those constructions. \\\\ 
  This paper is structured as follows: In Section~\ref{sec:Summary} we summarize our main mathematical result and sketch the argument. The full proof is deferred to Section~\ref{sec:5} and aims at the mathematically inclined reader. In Section~\ref{sec:overview} we give a pedagogical review of congruence subgroups, modular curves and their connection to torsion points of elliptic curves. This allows us in Section~\ref{sec:mcurves} to interpret the presence of singular fibers in torsion models directly from properties of the modular curves.  
    In Section~\ref{sec:4} we interpret our result in terms of  {\it swampland} constraint on the order of non-simply connected groups within F-theory. We close with with a summary and outlook in Section~\ref{sec:6}.

\section{Summary of Main Arguments}
\label{sec:Summary}

Our main result is: \\\\
{\bf Theorem 5.12} 
Let $\pi : Y \rightarrow B$ be a smooth   elliptically fibered  Calabi-Yau 3-fold with non-constant $j$-invariant.
Then $MW(Y/B)_{tors}$ is one of the following groups: 
\begin{align}
\begin{split}
\label{eq:torsionmodels}
  \ZZ_n:& \qquad (n = 1,2,3,4,5,6)\, ,   \\
  \ZZ_2 \times \ZZ_{2m} :& \qquad (m = 1,2)\, , \qquad \ZZ_3 \times \ZZ_3 \,   .
  \end{split}
\end{align}
We give a sketch of the proof in this section, and defer the details to Section~\ref{sec:5}.

\begin{enumerate}

\item{To begin, we use the Calabi-Yau condition to limit the possibilities for $B$.  
By the canonical bundle formula for elliptic fibrations,
$\om_Y \equiv \pi^*(\Ll_{X/B} \otimes \om_B)$.
In order for $Y$ to have trivial canonical bundle, either $\Ll_{X/B}$ and $\om_B$ are both trivial or $B$ is Fano.
If $\Ll_{X/B}$ is trivial, then the Weierstrass coefficients are constants, so the fibration is isotrivial, a contradiction.
Thus, $B$ is necessarily Fano.}

\item{By the classification of surfaces, $B$ is necessarily rational.
To limit the number of bases we need to consider, we replace $Y\rightarrow B$ by a birational model $Y_0 \rightarrow B_0$ obtained by contracting all exceptional curves on $B$.
This allows us to restrict attention to singular elliptic 3-folds over minimal rational surfaces that never reaches $(8,12,24)$ vanishing of $(f,g,\Delta)$ over a codimension 2 locus. The bases we have to consider are $\PP^2$ and the Hirzebruch surfaces $\FF_n$ for $n = 0, 1, \hdots, 12$.}

\item{Let $\eta\subset B$ be the generic point and $E/\eta$ be the fiber over the generic point.  
If we assume that the fibration has an $n$-torsion section (or a large torsion group of product type),
then there is a rational map $B \rightarrow C$ to a modular curve, and the coefficients of $E$ can be obtained by pulling back the Weierstrass coefficients of a universal elliptic curve via the map $B_0\rightarrow C$ . In other words, we have a diagram: 
\begin{equation}
\begin{tikzcd}
Y \arrow[r, dashed, "\Phi"] \arrow{d}{\pi}
&S_2 \arrow{d}{p}\\
B \arrow[r,dashed,  "\phi"]  &C_1  \, ,
\end{tikzcd}
\end{equation} 
where $C\cong X_1(n)$ and $S\rightarrow C$ is the N\'eron model for the universal elliptic curve with an $n$-torsion section. 
}

\item{To prove our main result,
we will show that:
\begin{itemize}
\item{The map $B\rightarrow C_1$ cannot be a morphism if $B$ is a minimal rational surface and $Y$ is Calabi-Yau.}
\item{For any $b \in B$ where $\phi$ is not defined, the Weierstrass coefficients over $b$ are guaranteed to vanish to order at least $(4d,6d,12d)$, where $d$ is an integer that depends only on $S\rightarrow C$.}
\end{itemize}
Finally, we show that $d = 1$ is only possible if the torsion group is one of the 10 groups in our list of \eqref{eq:torsionmodels}. 
Specifically, we will show that the degree of $\Ll_{S/C}$ divides $d$,
and for large enough torsion groups,
the minimal singularities imposed by the cusps on the modular curve force the degree of $\Ll_{S/C}$ to exceed 1.}
For $d>1$ it is known, that the total space does not admit a crepant resolution. We note that the singularities aforementioned singularities also appear over codimension 2 loci in elliptically fibered Calabi-Yau 4-folds and higher.  
\end{enumerate}  
 
In Appendix~\ref{app:ellipticsurfaces} , we show explicitly how to construct the map $\phi : B\rightarrow X_1(n)$ from the Weierstrass equation of an elliptic fibration with a chosen $n$-torsion section. In Section~\ref{ssec:Fintro} we will then study the application of our main theorem within the physics of F-theory.

\section{Modular curves}
\label{sec:overview} 

Modular curves come up multiple times in this paper  and are not standard in the physics literature,
so we start by giving a brief review of the theory. \footnote{For a more detailed exposition, see \cite{diamondshurman}.}
In this section, we will describe the geometry of modular curves viewed as analytic objects.
In the appendix, we also give an algebraic construction of general modular curves, together with equations for associated universal elliptic curves which are more useful in proving our main result.
The analytic perspective, described in this section is helpful in explaining the minimal configurations of singularities observed on elliptic fibrations with prescribed torsion.

\subsection{Review of Modular Curves}

\subsubsection*{"The" Modular Curve} 
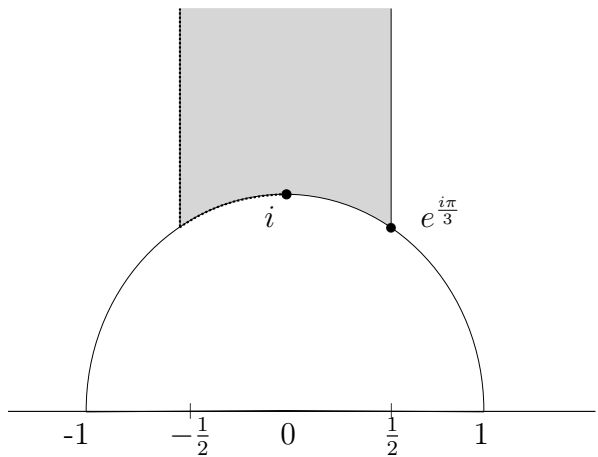
\begin{figure}
\centering
\begin{picture}(0,140)
\put(-100,20){
\includegraphics[scale=0.5]{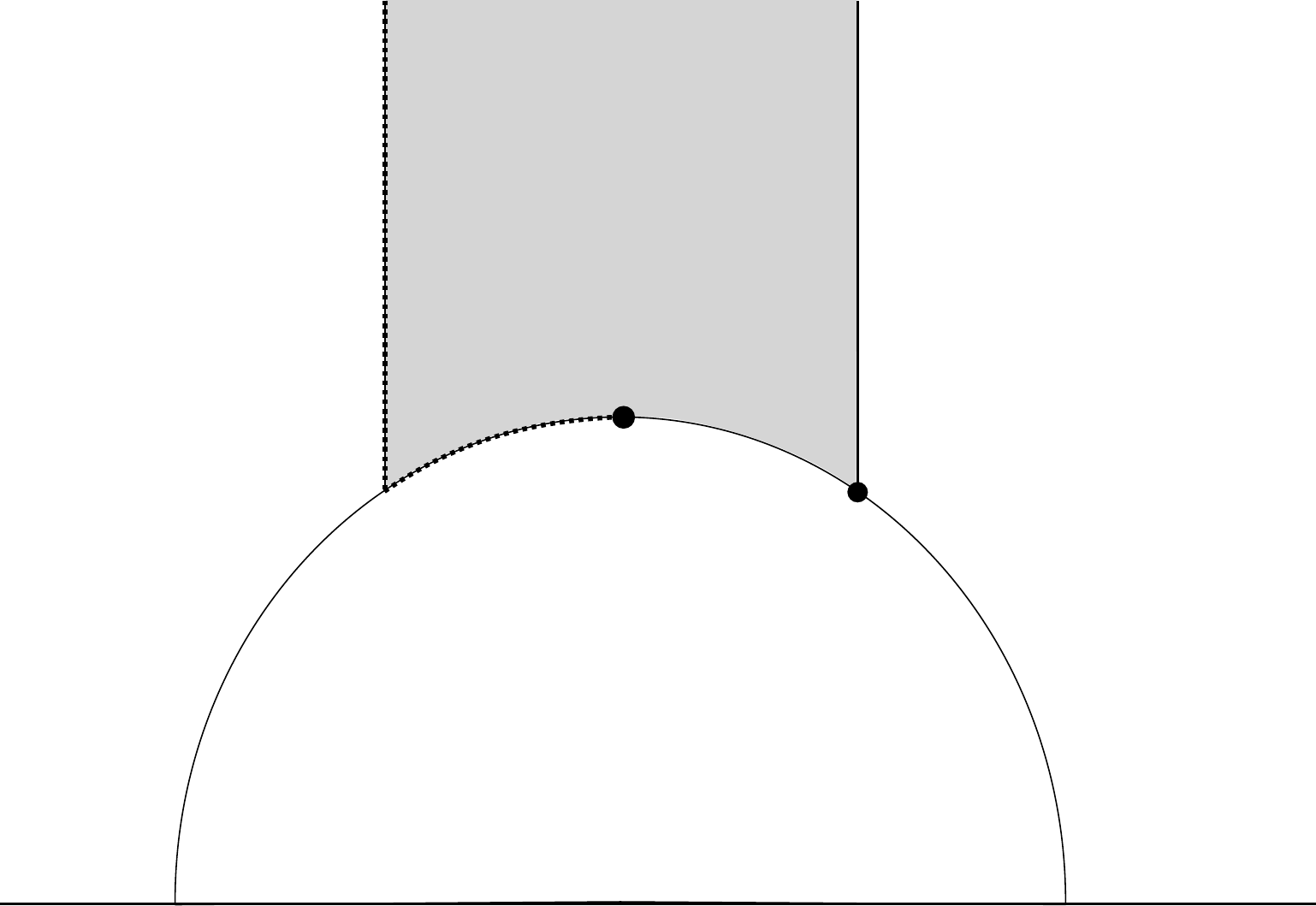}}
\put(-75,8){-1}
\put(80,8){1}
\put(7,8){0}
\put(48,19){{\tiny |}}
\put(-32,19){ {\tiny |}}
\put(-35,8){$-\frac12$}
\put(46,8){$\frac12$}
\put(,90){$i$}
\put(60,90){$e^{ \frac{i \pi }{3}}$}

\end{picture}
\caption{\label{fig:HalfPlane}{\it  The Fundamental domain of $\tau$ for $X(1) = \mathcal{H}/\mathrm{SL}(2,\ZZ)$.}}
\end{figure}

The modular curve $X(1)^o$ parametrizes isomorphism classes of elliptic curves $\CC$.
\begin{itemize}
\item{Algebraically, every elliptic curve admits a short Weierstrass equation:
\begin{align}
y^2 = x^3 + fx + g \, , \qquad (f,g \in \CC, 4f^3+27g^2 \neq 0) \, .	
\end{align}
Two such elliptic curves, with coefficients $f_i, g_i$, $i= 1,2$,
 are isomorphic if and only if there exists $\lambda \in \CC^\times$
such that $f_1 = \lambda^4 f_2$ and $g_1 = \lambda^4 g_2$.

Thus, $X(1)^o$ can be identified with the quotient:
\begin{align} X(1)^o = \set{(f,g) \in \CC^2 : 4f^3 +27g^2 \neq 0} /\left((f,g) \sim(\lambda^4 f, \lambda^6 g)\right) \, . \end{align}
Note that this quotient has two singular points,
namely the orbits $[(1,0)],[(0,1)]$, corresponding to the two elliptic curves with complex multiplication.
}

\item{Analytically,
Riemann's uniformization theorem shows that every elliptic curve is isomorphic,
as a Riemann surface, to $\CC/\Lambda$ , where $\Lambda$ is the lattice of periods of the elliptic curve.
Two quotients give rise to isomorphic elliptic curves if and only if the corresponding lattices are homothetic, i.e. scalar multiples of one another.

To compute the moduli space, we assume the lattice has been scaled so that one of the basis vectors is at 1 and the other basis vector lies in the (open) upper half plane, depicted in Figure~\ref{fig:HalfPlane}.
This allows us to identify points $\tau \in \Hh$ with isomorphism classes of elliptic curves: the point $\tau$ represents the elliptic curve $\CC/(\ZZ\oplus \ZZ \tau)$.
Two points $\tau, \tau'\in \Hh$ represent the same elliptic curve if and only if they are in the same $\mathrm{SL}(2,\ZZ)$-orbit (here $\mathrm{SL}(2,\ZZ)$ is acting on $\Hh$ by fractional linear transformations).
The region: 
\begin{align}
 \mathcal{D} = \set{ \tau \in \Hh : 1< |\tau | , -\frac{1}{2} < Re(\tau) \leq \frac{1}{2} } \cup \set{e^{i\theta} : \frac{\pi}{3} \leq \theta \leq \frac{\pi}{2}} \, , \end{align}
is a fundamental domain for the quotient,
so points in $\mathcal{D}$ are in bijection with points in $X(1)^o$.
The points $i$ and $e^{   i \pi /3}$  are cone points on $\Hh/\mathrm{SL}(2,\ZZ)$,
and correspond to the aforementioned singular orbits  $(1,0)$ and $(0,1)$ of $(f,g)$.
They represent the square and hexagonal lattices, which have extra symmetries.}

\end{itemize} 
The celebrated $j$-function, which is defined using Eisenstein series in the analytic setting or simply as:
\begin{align}
 j([(f,g)]) = 1728 \cdot \frac{4f^3}{4f^3+27g^2}\, , \end{align}
 in the algebraic one,
 gives a bijection between $X(1)^o$ and $\AA^1$. 

We can compactify $X(1)^o$ by adding a single point to the moduli space representing
the isomorphism class of an $I_1$ curve.
Analytically, we achieve this by taking the quotient of the extended upper half plane $\Hh^* = \Hh \cup \QQ \cup \set{\infty}$.
The compactified modular curve is denoted $X(1)$, and is isomorphic to $\PP^1$ as a Riemann surface.

\subsection{General Modular Curves}

Our goal now is to understand the moduli space of pairs $(E, p)$ where $E$ is an elliptic curve over $\CC$ and $p$ is a point of order exactly $n$ on $E$.

\begin{itemize}
\item{Every elliptic curve over $\CC$ is isomorphic to $E_\tau$ for some $\tau \in \Hh$.}
\item{Every point of order $n$ on $E_\tau$ has the form $\frac{c\tau + d}{n}+\Lambda_\tau$ for some pair of integers $c,d$ satisfying $\gcd(c,d,n) = 1$.}
\end{itemize} 
Thus, it suffices to study pairs $(E_\tau, \frac{c\tau + d}{n})$ for $\tau \in\Hh$ and $c,d\in \ZZ$ satisfying $\gcd(c,d,n) = 1$.
We now need to determine when two pairs $(E_\tau, \frac{c\tau + d}{n}), (E_{\tau'}, \frac{c'\tau' + d'}{n})$ are isomorphic.
We will show that the action of $\mathrm{SL}(2,\ZZ)$ on the space of bases of $\Lambda_\tau$ induces an action on the $n$-torsion points of $E_\tau$. \\
Let $p \in E_\tau$ a point on the elliptic curve, that we can write uniquely as 
 \begin{align}p:\quad   x\tau + y + \Lambda_\tau \, \text{ for }\, x,y \in [0,1)\, . \end{align} 
For $\gamma = \ttm{a}{b}{c}{d} \in \mathrm{SL}(2,\ZZ)$, we define the action on that point $p$ as 
\begin{align} \gamma \cdot p = (a\tau + b) x + (c\tau + d) y + \Lambda_\tau \, . \end{align} 
It's clear that $\gamma$ induces an automorphism of $E_\tau$ as a group,
so it necessarily restricts to an automorphism of the $n$-torsion of $E_\tau$.
Thus $\mathrm{SL}(2,\ZZ)$ acts on $n$-torsion pairs $(E_\tau, \frac{c\tau+d}{n})$.
We next show that it acts transitively on such pairs:
in other words for any pair $(E_\tau, p + \Lambda_\tau)$,
we can find $\gamma \in \mathrm{SL}(2,\ZZ)$ such that $\gamma \cdot p + \Lambda_\tau = \frac{1}{n} + \Lambda_\tau$.
This will show that the moduli space of pairs $(E,p+\Lambda_\tau)$ is isomorphic to $\Hh/\Gamma_1(n)$, where $\Gamma_1(n)$ is the stabilizer of the pair $(E_\tau, \frac{1}{n} + \Lambda_\tau)$.  
Let $\frac{c\tau.+ d}{n} + \Lambda_\tau$ be an arbitrary point of order $n$.
Since $\gcd(c,d,n) = 1$,
there exist integers $a,b,k$ such that $ad-bc + kn = 1$.
That is equivalent to saying there is a matrix in $\mathrm{SL}(2,\ZZ_n)$ with entries congruent to $\gamma$ mod $n$.
The reduction map $\mathrm{SL}(2,\ZZ)\rightarrow \mathrm{SL}(2,\ZZ_n)$ is surjective,
so there exists $\gamma \in \mathrm{SL}(2,\ZZ)$ such that $\gamma \cong \ttm{a}{b}{c}{d} \pmod n$. 
Observe now the inverse action of $\gamma$ on the pair
\begin{align}
\gamma\inv \cdot \left(E_\tau, \frac{c\tau + d}{n}+\Lambda_\tau\right) = \left(E_\tau, \frac{c(d\tau - b)+d(-c\tau + a)}{n} +\Lambda_\tau\right) = \left(E_\tau, \frac{1}{n}+\Lambda_\tau\right) \, .
\end{align}
Thus, if we take a larger fundamental domain in $\Hh$,
we do not have to keep track of the specific coefficients $c,d$ for the point of order $n$.
The problem is now reduced to understanding the stabilizer of $(E_\tau, \frac{1}{n}+\Lambda_\tau)$.\\
Specifically, we want a characterization of those $\gamma \in \mathrm{SL}(2,\ZZ)$ that fix the torsion point
\begin{align}
\gamma \cdot \frac{1}{n}+\Lambda_\tau = \frac{1}{n}+\Lambda_\tau \, . \end{align}
\begin{figure}
\begin{picture}(0,100) 
\put(70,8){\includegraphics[width=1.8in]{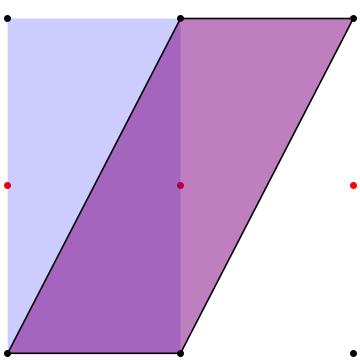}}
\put(270,8){
	\includegraphics[width=1.8in]{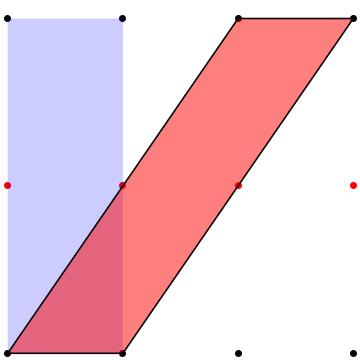}}
	
\put(50,110){$a.)$}	
\put(250,110){$b.)$}	
\end{picture}   
\caption{\label{fig:g12}{\it   Above we have  chosen  a fundamental domain colored in blue for $E_\tau$ the depiction of a 2-torsion point $\frac{\tau}{2}+\Lambda_\tau$ in orange. We act on the basis by the generator $T$ in $a.)$ 
which translates the point $\frac{\tau}{2} + \Lambda$ to $\frac{\tau+1}{2} + \Lambda$. In $b.)$ we act on the fundamental domain by $T^2$ in to obtain the pink one while fixing the torsion point.}}
\end{figure} 
The new basis is given by  $a\tau + b, c\tau + d$,
and $\frac{1}{n}+\Lambda_\tau$ is being mapped to $\frac{c}{n} \tau + \frac{d}{n} \Lambda_\tau$.
It is clear that $\frac{1}{n}$ is fixed exactly when $\frac{c}{n} \in \ZZ$ and $\frac{d}{n} \in \frac{1}{n} + \ZZ$.  An illustration of this action on the torus lattice is depicted in Figure~\ref{fig:g12} for a chosen order two torsion point. 
In particular, we need the entries to satisfy
\begin{align}
c \equiv 0 \pmod n \quad \text{ and } \quad d \equiv 1 \pmod n \, .
\end{align} 
It then follows, that the matrices need to satisfy
\begin{align}
\ttm{a}{b}{c}{d} \equiv \ttm{*}{*}{0}{1}\pmod n \,\, \, \equiv \ttm{1}{*}{0}{1} \pmod n \, ,
\end{align} 
where we have used $ad - bc = 1$ and $c \equiv 0 \pmod n$ such that we   must have $ad \equiv a \equiv 1 \pmod  n$. \\\\
To summarize, a matrix $\ttm{a}{b}{c}{d} \in \mathrm{SL}(2,\ZZ)$ fixes $(E_\tau, \frac{1}{n} +\Lambda_\tau)$ as a torsion pair if and only if 
 \begin{align*}\ttm{a}{b}{c}{d} \equiv \ttm{1}{*}{0}{1} \pmod n \, . \end{align*}
We thus define:
\begin{align}
\label{eq:gamma1}
\Gamma_1(n) &= \set{\gamma \in \mathrm{SL}(2,\ZZ) : \gamma \equiv \ttm{1}{*}{0}{1} \pmod n} \, .
\end{align}
We can then go ahead and define the (open) modular curve as the quotient $X_1(n)^o = \Hh/\Gamma_1(n)$.
\footnote{This notation is not standard. In the modular curves literature, the curve we are denoting as $X_1(n)^o$ is usually denoted $Y_1(n)$.
We've deviated from the standard notation to avoid confusion with the Calabi-Yau total spaces, which we are denoting $Y$.} 
We identify points on $X_1(n)^o$ with pairs consisting of an elliptic curve $E_\tau$ and a point of order $\frac{c\tau + d}{n}$.
The compactified modular curve $X_1(n)$ is obtained by taking the quotient of the extended upper half plane by $\Gamma_1(n)$.
The cusps of the modular curve are the points in $X_1(n)\backslash X_1(n)^o$.

There are also other examples of modular curves that appear frequently in the literature.
These are denoted $X(n)$ and $X_0(n)$, and are defined as subgroups of the upper half-plane by the subgroups:
\begin{align}
\label{eq:gamma}
\Gamma(n) &= \set{ \gamma \in \mathrm{SL}(2,\ZZ) : \gamma \equiv \ttm{1}{0}{0}{1} \pmod n} \, ,\\
\Gamma_0(n) &= \set{ \gamma \in \mathrm{SL}(2,\ZZ) : \gamma \equiv \ttm{*}{*}{0}{*} \pmod n} \, .
\end{align} 
As moduli spaces, $X(n)$ classifies pairs $(E,(p_1,p_2))$ where $p_1, p_2$ are a ``nice" basis of the $n$-torsion of $E$, and $X_0(n)$ classifies pairs $(E, \ip{p})$ where $\ip{p}$ is a cyclic subgroup of order $n$.

\subsection{Cusps}
 
Finally, we need to define the width of a cusp, as this idea plays an important role to discuss minimal singularities in Section~\ref{sec:mcurves}.
 To that end, we introduce the following notation.
Set:
\begin{align}
\mathrm{SL}(2,\ZZ)_\infty &= \set{\ttm{1}{n}{0}{1}: n \in \ZZ}
\end{align}
This is the stabilizer subgroup of $\infty$ for the action on $\Hh^*$ by fractional linear transformations. 

For a general subgroup $\Gamma \subset \mathrm{SL}(2,\ZZ)$, we define:
\begin{align}
\Gamma_\infty = \Gamma \cap SL(2,\ZZ)_\infty =  \set{ \gamma \in \Gamma : \gamma = \ttm{1}{a}{0}{1} }
\end{align} 

Now, fix a congruence subgroup $\Gamma$.
Each cusp of the associated modular curve corresponds to the orbit of some $s \in \QQ$ under $\Gamma$.
Choose such an $s$, and choose $\delta \in \mathrm{SL}(2,\ZZ)$ such that $\delta \cdot s =\infty$.
The {\bf width of the cusp $x$}, denoted $h(x)$, is abstractly defined as:
\begin{align}
h(x) = [\mathrm{SL}(2,\ZZ)_\infty : (\delta \set{\pm I} \Gamma \delta\inv)_\infty ] \, .
\end{align}
This number encodes the smallest integer $h$ such that $\ttm{1}{h}{0}{1} \in \delta \Gamma \delta\inv$;
geometrically, we have acted on the upper half plane to put $s$ at the ``top" of the picture,
and we're counting the number of vertical strips in the fundamental domain.
We can also compute the width using the triangulation we pull back from $X(1)$,
by dividing the number of triangles that meet at a given cusp by 2. We will discuss several concrete examples in the following sections. 

From the moduli perspective, the width of the cusp encodes the ramification over the $I_1$ point on $X(1)$, and thus determines the minimal singularities on the N\'eron model of the universal elliptic curve.
Specifically, whenever there is a cusp of width $h$ on the modular curve,
the associated elliptic surface should have an $I_h$ singularity (or worse).\footnote{These points must be of $I_h$ type, since $f$ and $g$ are non-vanishing, as they come from special points in the fundamental domain of $X(1)$, that are not mapped onto the real axis.}
This will be discussed further in the minimal singularities section.

\subsection{Minimal Singularities}
\label{sec:mcurves}
When considering explicit Weierstrass models we find (e.g. see Section~\ref{sec:4}) that
  the presence of torsion in $MW$ imposes a minimal number of singular fibers. In the context of F-theory this is surprising, as it forces a minimal gauge group when the full fibration admits torsion. This is particularly true for larger order torsion, where not one but multiple singular fibers/gauge algebra factors are introduced. 
 For example, any fibration with a section of order 5 necessarily has at least two fibers of type $I_5$ (or possibly $I_{5d}$ for some integer $d$). 
For fibrations with 7-torsion, the number of $I_{7d}$ fibers has to be at least 3.
The exact configuration of fibers which is imposed can be computed from the Weierstrass equation of the universal elliptic curve (e.g. see Appendix \eqref{app:ellipticsurfaces}), but obtaining such equations can often be tedious.

Fortunately, the degenerate fibers and the Euler characteristic of the universal elliptic surface $S\rightarrow C$ can both be read off directly from the associated congruence subgroup.
Geometrically, the singularities appear because of the cusps on the compactified modular curve. 
The information needed to determine which singular fibers appear in $S\rightarrow C$ is encoded in the cusps of the modular curve:
\begin{itemize}
\item{The points on the discriminant locus are in bijection with cusps on the compactified modular curve.}
\item{The fiber over a point in the discriminant locus is of type $I_d$, where $d$ is the width of the associated cusp.}
\end{itemize}
Both the number of cusps and their widths can easily be computed using the standard triangulations of the modular curves:
the cusps are the points on the real line, and the width of each cusp can be computed by dividing the number of triangles meeting at the cusp by 2.
They can also be computed using algebraic methods,
see e.g. \cite{diamondshurman}.

The number of triangles in the triangulation is equal to twice the sum of the widths.
As a result, it can be interpreted as twice the degree of the discriminant of an elliptic fibration with only the minimal singularities forced by the modular curve.
The degree of the discriminant is useful, because it determines the degree of the fundamental line bundle of the universal surface, which is crucial to our analysis.

Since the triangulation is pulled back from the map $X\rightarrow X(1)$,
the number of triangles is exactly the degree of the map,
which is the index $[\mathrm{SL}(2,\ZZ):\Gamma]$.

In Appendix~\ref{app:indices}, we derive formulas for the index of all congruence subgroups.
For example, the index of $\Gamma_1(n)$ is:
\begin{align}
 [\Gamma(1) :\Gamma_1(n)] = n^2 \prod_{p | n} 1 - \frac{1}{p^2}\, , \end{align}
where the product is taken over all primes dividing $n$.
One can check (see Appendix~\ref{app:indices}) that this index is equal to 12 when $n = 4$ and is divisible by 24 for all $n > 4$.
For $n > 4$, the index can be used to compute the degree of $\Ll_{S/C}$,
where $S\rightarrow C$ is the N\'eron model of the universal elliptic curve:
\begin{align}
\label{eq:gamma1}
[\Gamma(1):\Gamma_1(n)] = 24 \deg \Ll_{S/C} \, .\end{align} 
Thus, when $C = Y_1(n)$, we have $\deg \Ll_{S/C} = 1$ for $n = 5,6$,
$\deg \Ll_{S/C} = 2$ for $n = 7,8$, $\deg \Ll_{S/C} = 3$ for $n = 9,10$, etc.
This matches up exactly with the tables in \cite{MP},
and confirms that the $\ZZ_5, \ZZ_6$ rational elliptic surfaces are extremal,
and similarly that the $\ZZ_7, \ZZ_8$ K3 surfaces are extremal, and any larger order has Kodaira dimension 1,  hence not Calabi-Yau.  
\begin{figure}
\begin{picture}(0,200)
\put(10,0){\includegraphics[width=3in]{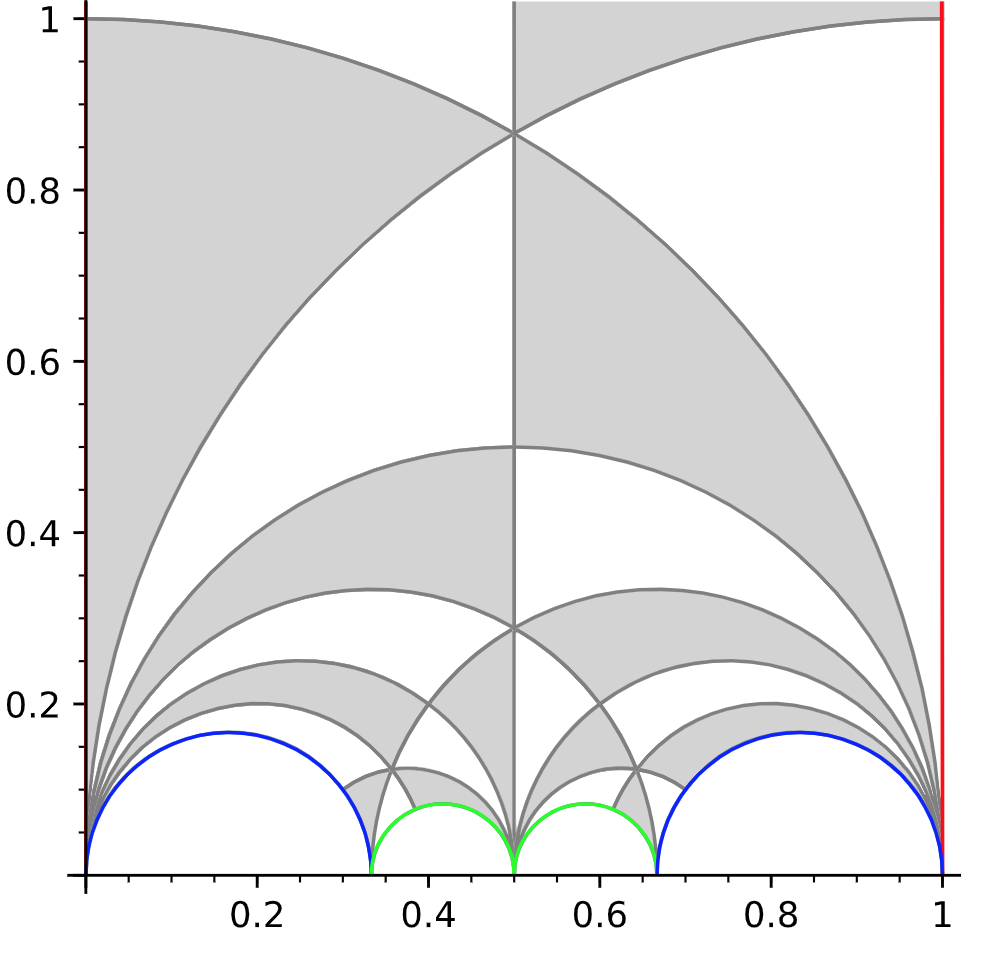}}
\put(250,10){
	\includegraphics[width=3in]{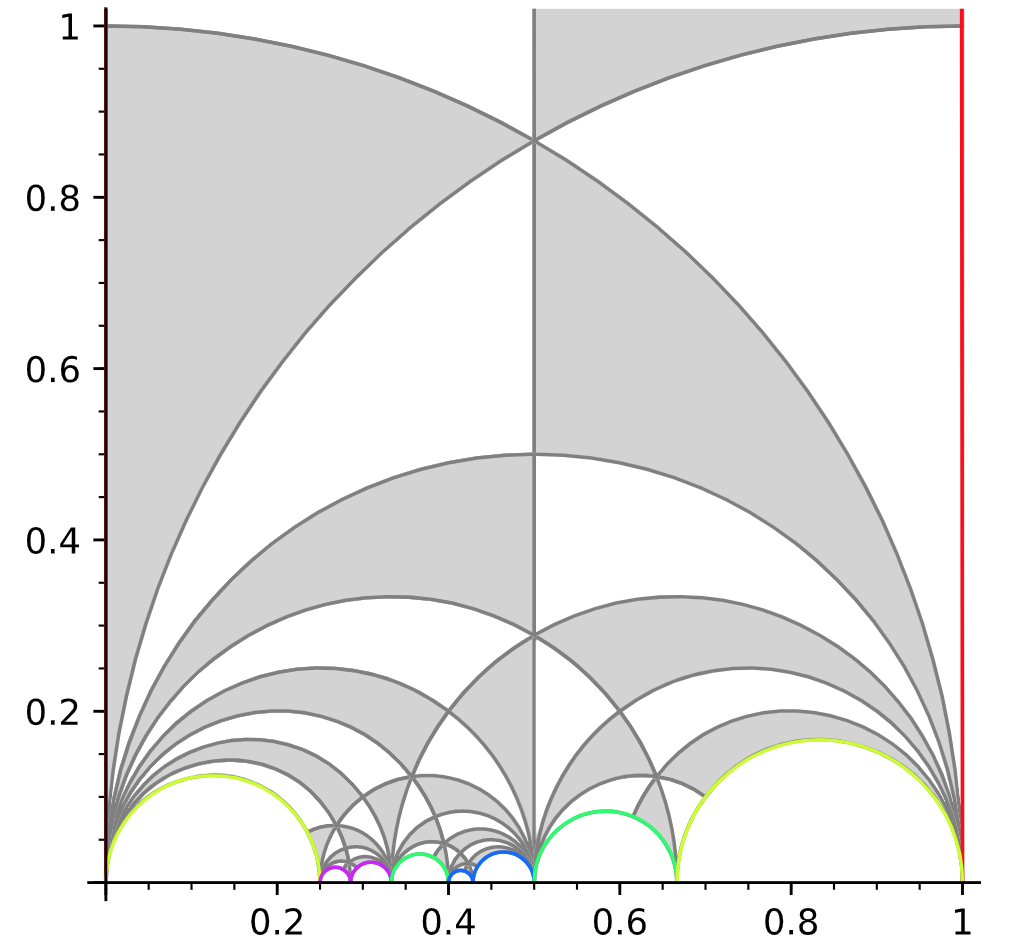}}

\end{picture}  
	\caption{\label{fig:X16}{\it  Triangulation of $X_1(6)$ and  $X_1(7)$. Sides with equal colors are to be identified.}}
\end{figure}
E.g. the triangulation for $X_1(6)$ and $X_1(7)$ are depicted in Figure~\ref{fig:X16} using {\it Sage} \cite{Sage,SageKurth} which admits the following cusp points
\begin{align}
S_{\text{cusp}}^{(6)}: \{ 0^{(6)}, 1/3^{(2)}, 1/2^{(3)}, \infty^{(1)} \}\, ,
\end{align}
with their respective widths denoted as superscripts.
This means the universal surface should have fibers of type $I_6, I_2, I_3, I_1$. These are exactly the discriminant loci, that we also find in the generic Weierstrass model that we discuss in Section~\ref{ssec:hightorsion}.

Similarly, for $X_1(7)$, we find that there are 6 cusps (including $\infty$) at
\begin{align}
S_{\text{cusp}}^{(7)}: \{ 0^{(7)}, 2/7^{(1)}, 1/3^{(7)}, 3/7^{(1)}, 1/2^{(7)}, \infty^{(1)} \}\, ,
\end{align} 
three of which have width 7 and three of width 1, so the universal surface has 3 $I_7$ fibers and 3 $I_1$ fibers. 
This explains why we can't construct a model with a 7-torsion section and with exactly one $I_7$, e.g but predicts the right fibers in the K3, (see Section~\ref{ssec:K3}).
   
The congruence subgroup that leaves two $\mathbb{Z}_n$ torsion points, is $\Gamma(n)$ as defined above. We compute the index in Appendix~\ref{app:indices}, and repeat the result here as
\begin{align}
\label{eq:gamma}
 [\Gamma(1) :\Gamma(n)] = n^3 \prod_{p | n} 1 - \frac{1}{p^2} \ . \end{align}
For $n=2,3,4,5$ the indices are computed to be $6,24,48,120$. With the identification of the fundamental line bundle analogous to Equation~\eqref{eq:gamma1}, this shows, that $\mathbb{Z}_3 \times \mathbb{Z}_3$ is the maximum allowed torsion possible in rational elliptic surfaces, three and fourfolds. The $\mathbb{Z}_4 \times \mathbb{Z}_4$ torsion extremizes K3 and is depicted in  Figure~\ref{fig:Z9} of Section~\ref{ssec:K3}, where we will come back to this topic. Higher order torsion points are not allowed. 
   
\subsection{Remarks about the $\ZZ_4$ Case}
\label{ssec:z4}
When $n \geq 5$, the singularities predicted by the widths of the cusps correspond exactly to the fibers observed on the universal surface.
This is not the case when $n = 4$. 
We wish to briefly address this peculiarity\footnote{There are other reasons to believe the $\ZZ_4$ group is special. For example, the group structure in the special fiber of an elliptic surface with an $I_{2k+1}^*$ is abstractly isomorphic to $\ZZ_4$, which means we can have non-semistable fibrations with 4 torsion.} in this section.

For $n > 4$, the sum of the widths adds up to an integer multiple of the Euler characteristic of the universal surface,
allowing us to deduce the degree of the fundamental line bundle using Equation~\eqref{eq:gamma1} and to determine the minimal singularities without needing a Weierstrass model. 
When $n = 4$, the index of the congruence subgroup is 12, so the formula breaks down. 
As a result, we see a discrepancy between the singularities predicted by the widths of the cusps observed on the modular curve and the singularities on the universal surface with a 4-torsion section:
the modular curve only forces $I_4+I_1+I_1$ on the elliptic curve,
but since the discriminant of any rational elliptic surface has degree 12, we need additional singularities to obtain a smooth elliptic surface over $\PP^1$.  

From the algebraic perspective, we have the equation for the universal elliptic curve:
\begin{align} y^2 + xy - t y = x^3 - tx^2 \, ,  \end{align}
which extends to:
\begin{align} y^2 + t_1^2xy - t_0t_1^2 y = x^3 - t_0t_1x^2 \, . \end{align}
Note that this equation has an $I_1^*$ singularity, which is the only type we see an additive fiber in an elliptic fibration with a torsion point of order at least 4.
After a quadratic base change, we can also obtain a semistable rational surface which has a 4-torsion point.
The short Weierstrass equations for the two models are given below:
\begin{align}
y^2 &= x^3 + t_1^2 \frac{16t_0^2 - 16 t_0t_1 +t_1^2}{48}x - t_1^3 \cdot \frac{64t_0^3 +120t_0^2t_1-24t_0t_1^2 +t_1^3}{864} \, ,\\
y^2 &=  x^3 + \frac{16t_0^4 - 16 t_0^2t_1^2 +t_1^4}{48}x - \frac{64t_0^6 +120t_0^4t_1^2-24t_0^2t_1^4 +t_1^6}{864} \, .
\end{align}
We will come back to the $\mathbb{Z}_4$ model in Section~\ref{ssec:hightorsion} from a more general perspective.  
\section{Bounds on Non-simply connected Gauge Groups from F-theory}
\label{sec:4} 
In this section we want to apply the bounds on torsion within the context of F-theory.  
Hence we first review F-theory and the role of Mordell-Weil torsion in its effective theory as first investigated in\cite{Aspinwall:1998xj} and further explored in \cite{Mayrhofer:2014opa} which can be skipped by any expert in the field.   
The bounds on torsion, which were discussed in the sections before and are proved  more rigorously  in Section~\ref{sec:5}  can be translated into bounds on non-simply connected gauge groups in SUGRA theories obtained from F-theory. As we show, those bounds are surprising  from pure 6-dimensional SUGRA arguments and their massless spectra. 
 Finally we also comment on bounds on smooth Calabi-Yau quotient torsors and their connection to discrete symmetries and superconformal matter.
\subsection{F-theory and the Role of Torsion}
\label{ssec:Fintro}
This section is intended as a recap of Mordell-Weil torsion in F-theory models, based on \cite{Mayrhofer:2014opa}.\\ In F-theory, one considers an elliptically fibered threefold:
\begin{align}
\begin{array}{cc}
\mathcal{E}  \rightarrow& Y_n\\
 & \downarrow \pi \\
 & B_{n-1} 
 \end{array} \, ,
\end{align} 
given by a Weierstrass equation:
\begin{align}
y^2 = x^3 + fx + g 
\end{align}
The local axio-dilaton $\tau = C_0 + i g_{IIB}^{-1}$ of type IIB string theory is identified with the complex modulus of each fiber, and is allowed to vary over the complex base $B_{n-1}$. 
 The singular fibers are located over the vanishing set of the discriminant:
 \begin{align}
 \Delta = 4f^3 + 27g^2 \, .
 \end{align}
  The codimension one components of the discriminant locus
  \begin{align}
\delta_i \subset V(\Delta)\, ,
  \end{align}
  have the interpretation of generalized stacks of [p,q]-7-branes that host a local gauge algebra $\mathcal{G}_i$ according to their Tate-fiber type. The order of $\Delta$ can be enhanced, e.g. over the collision of components at codimension two $\delta_{i,j}=\delta_i \cap \delta_j$, which gives rise to matter multiplets that form representations $\mathbf{R}$ of the algebra $\mathcal{G}$. The matter representation can be inferred from the enhancement to the local algebra $\mathcal{G}_{i,j}$ via the decomposition \cite{Katz:1996xe} as: 
 \begin{align}
 \mathbf{R}_{i,j} = adj(\mathcal{G}_{i,j})/ (adj (\mathcal{G}_i) \oplus adj (\mathcal{G}_j))  \,.
 \end{align}
This construction gives a powerful tool to construct and  classify a large class of consistent $D=10-2d$ dimensional SUGRA theories via geometric methods.   \\
The physics of F-theory compactifications on CY $d$-folds $Y_d$ with non-trivial MW torsion group, say MW$_{tor}(Y_d)=\mathbb{Z}_n$, is understood as a global gauging of the center of the gauge algebra $\mathcal{G}_i$ resulting in a non-simply connected  gauge group $G$:
 \begin{align}
 G = \frac{\prod_i \mathcal{G}_i }{\mathbb{Z}_n} \times \hat{G} \, , \qquad \pi_1 (G) = \mathbb{Z}_n \, .
 \end{align}
 The quotient factor induces a  constraint matter spectrum by projecting out representations that have a non-trivial center charge $q_{cen}$. 
In general, the restriction to center charges can be seen explicitly via the construction of the {\it torsion Shioda map} $\Sigma$ in the resolved geometry. For any non-trivial torsion point $S_k$ and zero-section $Z$, the Shioda map defines a map from the elliptic curve to the N$\acute{\text{e}}$ron-Severi lattice of   $Y_d$ \footnote{We ignore contributions from vertical divisors at this point.}. As we have argued in the sections above, the mere presence of torsion points $S_k$, enforces the presence of singularities that need to be resolved leading to the following form of the Shioda map:
\begin{align}
\Sigma(S_k) = S_k-Z + (S \cdot c_i) C^{-1}_{i,j} f_j =S-Z + \frac1n \sum_i a_j f_j \, .
\end{align}
The fractional linear contribution of the resolution divisors $f_i$ above is determined by the $i$-th fibral curve $c_i$ in the resolved geometry, that is intersected by the torsion section $S$. The overall normalization $n$ is inherited from the inverse determinant of the Cartan matrix $C$, which is itself of the order of its center.
Coming from a torsion section $S_k$ implies the Shioda map to be a trivial divisor in $H^2(Y)$ which therefore can be rewritten as
\begin{align}
\Xi(S_k) = S_k - Z = -  \frac1n \sum_i a_j f_j \, , 
\end{align}
and represents an $n$-torsion element in the quotient cohomology of $H^{1,1}(Y)/\langle [f_i] \rangle_\mathbb{Z}$.
  At the intersection $\delta_{i,j}$, the fibral curves in the resolved model generically split further $c_i \rightarrow c_{i,m}$ reflecting the nature of the enhanced singularity. The exact representation $\mathbf{R}_{i,j}$ and compatibility with the gauged $\mathbb{Z}_n$ centers can be inferred from intersections with the resolution divisors $f_i$ and the fractional element of the torsion Shioda map:
\begin{align}
\mu_r=(c_{i,m}\cdot f_r  ) \, \qquad q_{cen}= c_{i,m}\cdot \Xi(S_k) \, .
\end{align}
All other weights can be obtained from adding fibral curves $c_i$ which also shows, that the center charge is $n$-fractional and only well defined modulo $1$. However, since the torsion Shioda map is a trivial divisor the center charge must be integral to be consistent with the global gauge group factor. Field theoretically one can understand this, by promoting the fractional piece of the torsion Shioda map to a  projector of the form $\Phi=\text{e}^{2\pi i \Xi(S_k)}$. This projector removes representations with the wrong (combination) of weights of the gauge group $G$.
Therefore the above element defines a generator of the n-refined (co-) weight lattice \footnote{Note that a similar effect can be resembled in the presence of a free Mordell-Weil group, that gives rise to a U(1) symmetry that can embed non trivially inside the center of other non-Abelian group factors \cite{Grimm:2015wda,Cvetic:2017epq}. Although having a similar origin, these factors can only be present in the presence of additional Abelian symmetries, which we do not want to focus on in this work.}. Note that the spectrum computed above, is just the massless one and therefore comprises only a small sub-sector of the full theory. However it seems convincing, due to the geometric realization of the torsion Shioda map, that the constrained (co-) charge lattice extends to the full massive sector of the theory as well. In more in general it is expected that the non-simply connected gauge groups $G$, also affects the presence of dyonic line and surface operators in the theory \cite{Aharony:2013hda}. Similarly we note, that the presence of torsion affects the interpretation of $(p,q)$-strings as combined states of fundamental strings and $D1$ branes. A general $(p,q)$-string (with p and q being co-prime) that couples to the IIB $(B_2,C_2)$-fields with the respective quanta can always be rotated into the fundamental $(1,0)$ string picture by virtue of an $\mathrm{SL}(2,\mathbb{Z} )$ transformation. However under the reduced congruence subgroups we do not have the full transformation group at our disposal and therefore not every $(p,q)$ combination allows for such a transformation.
As an example, for a parametrization of a $\Gamma_1(n)$ matrix, one may consider the following action on a fundamental string:
\begin{align}
\left( \begin{array}{cc}  1+ a\, n & r \\ k\, n & 1+b\,n  \end{array} \right)\left( \begin{array}{c } 1 \\ 0   \end{array} \right) = \left( \begin{array}{c } 1+a \, n  \\  k  \, n   \end{array} \right) \, , \qquad \text{ with } a,b,r,k  \in \mathbb{Z} \, .
\end{align}
One finds the  $D1$  brane charge has to be divisible by the order $n$ to allow for a rotation into the fundamental string picture. This might be interpreted in two ways: either the $(p,q)$ string states with odd $q$-$D1$ charge are fully absent from the spectrum, or these states can be present but come from a different sector of the theory which can not be described as a fundamental string in any given $\Gamma_1(n)$ frame. However, we take the non-trivial behavior of more general string states as evidence, that torsion also affects the massive sector of the theory in a non-trivial way and  would like to return to this interesting question in future work.
\subsection*{An $su(2)/\mathbb{Z}_2 $ example} 
\begin{figure} 
\centering
 \includegraphics[width=3in]{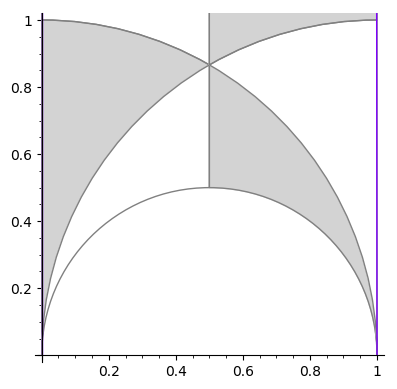} 
 \caption{{\it \label{fig:Z2}Triangulation of $X_1(2)$. A cusp point of width two is found at the origin.}}
\end{figure}

 The simplest direct example is that of an $su(2)/\mathbb{Z}_2 \sim SO(3)$ group. As SO(3) does not admit two dimensional representations, it is directly evident, that the fundamental of the $su(2)$ covering algebra, must be absent from the spectrum. Let's now engineer this model in F-theory, via an $\mathbb{Z}_2$ torsion point. There, the $su(2)$ gauge algebra is in fact forced upon us directly, as can be seen using the modular curve of $X_1(2)$. Explicitly, the $\Gamma_1(2)$ congruence subgroup is generated by upper triangle matrices modulo two. To find the generic singularity(s) we have to find the cusp(s) of the modular curve and their widths. As reviewed in Section~\ref{sec:mcurves}, we first have to find the triangulation of $X_1(2)$ by pulling back the standard triangulation of $X(1)$. The degree of this map is given in Equation~\eqref{eq:gamma1}  for $n=2$ which we explicitly compute to be:
\begin{align}
 [\Gamma(1): \Gamma_1(2)]=3 \, .
\end{align}
We can act on the fundamental domain of $X(1)$ by the coset representatives:
\begin{align}
\alpha_1: \left( \begin{array}{cc} 1 & 0 \\ 0 & 1\end{array} \right)\, , \quad \alpha_2: \left( \begin{array}{cc} 0 & -1 \\ 1 & 0\end{array} \right)\, ,\quad \alpha_3: \left( \begin{array}{cc}0 & -1 \\ 1 & 1\end{array} \right)\, ,
\end{align}
to obtain a fundamental domain for $X_1(2)$.
The triangulation of the modular curve into $2 \cdot 3$ \footnote{The number of subregions is the number of regions in the triangulation of $X(1)$ multiplied by the degree of the map.
The triangulation of $X(1)$ has 2 regions, and the degree of the map is equal to the index of the subgroup.} sub-regions  is explicitly shown in Figure~\ref{fig:Z2}. There are two cusps at the points:
\begin{align}
S^{(2)}_{cusp}: \{ 0^{(2)}, \infty^{(1)}   \}
\end{align}
with the one at the origin having width two. 
   This is consistent with the requirement to have the appropriate gauge algebra $\mathcal{G}$ present on which the $\mathbb{Z}_2$ quotient can act. 
We can now turn to the explicit Weierstrass model, which exactly reflects what we anticipated from the modular curve:
\begin{align}
\label{eq:WSFZ2}
y^2 =& x(x^2 + a_2 x + a_4) \, , \qquad a_4 \in \mathcal{O}(K_b^{-4})\, , \quad a_2 \in \mathcal{O}(K_b^{-2})\, . \\
f=& a_4 - \frac13 a_2^2 \, , \quad g= \frac{1}{27} a_2 (2 a_2^2 -9 a_4)\, , \quad
\Delta=  a_4^2 (4 a_4 - a_2^2) \, .
\end{align}
There we find the expected $I_2$ locus over $a_4=0$ and the residual $I_1$ locus. The torsion factor restricts the algebra to $su(2)/\mathbb{Z}_2 \sim SO(3)$. Hence, geometrically we expect the enhancement from the $I_2$ to an $I_3$ locus to be forbidden, simply because of the incompatible center. Instead the collision point we find is of type $III$, that has an associated $su(2)$ algebra as well, compatible with the correct center but giving rise to no new matter multiplets. In the resolved geometry (e.g. see \cite{Mayrhofer:2014opa}) it is possible to obtain the torsion Shioda map $\Sigma(S_1)$ (modulo vertical divisors) as:
\begin{align}
\Sigma(S_1)= S_1-Z + \frac12 f_1  \, ,
\end{align}
with $f_1$ being the $su(2)$ resolution divisor. By abuse of notation, we denote with $f_1$ also the possible $su(2)$ weight of some representation. At codimension two, any reducible matte curve must have vanishing (mod 1) intersection with $\Sigma(S_1)$. This shows, that every representation with odd $su(2)$ Cartan charge must be absent from the spectrum.\\\\ 
 Starting from any torsion model is a great way to systematically study and classify SUGRA theories in various dimensions with a non-trivial fundamental group. The Weierstrass model with the respective torsion point will in turn only allow for the factorization of ADE singularities that have a compatible center. This perspective is confirmed from the perspective of the congruence subgroups, that must be compatible with the local SL$(2,\mathbb{Z})$ monodromy of the singularity. We have summarized the Kodaira fiber types and their congruence subgroups\footnote{Compatible $\Gamma_0(n)$ subgroups are considered in \cite{Berglund:1998va}.} in Table~\ref{tab:fibermono}. 
In order to compare a fiber monodromy matrix $M$ to be compatible with a $\Gamma_1(n)$ or $\Gamma(n)$ congruence subgroup, requires us to bring it into a compatible form via an $g\in$ SL$(2,\mathbb{Z})$ conjugation 
\begin{align}
\widetilde{M}=g  \cdot M \cdot g^{-1} \, .
 \end{align} 
In practice this can be achieved by a combination of the matrices $g_1 = \left(\begin{array}{cc} 0 & 1 \\ 1 & 1 \end{array}\right)$ and the $S$ and $T$ generator. E.g. for an type $I^*_{2n+1}$ fiber we can use $g=  g_1^2$ to obtain
\begin{align}
\widetilde{M}_{I*_{2n+1}} = \left( \begin{array}{cc}1 +4n  & -1-2n\\ 2n+8n  & -3-4n \end{array} \right) \in \Gamma_1 (4)\in \Gamma_1(2) \, .
\end{align}
For the monodromies of type $I_n, III,IV^*$ and $III^*$ fibers, conjugation by $g=S\cdot g_1$. For type $IV$ and $I^*_{2n }$ one can conjugation by $g=S\cdot T \cdot S$  and  $g=g_1$ respectively to achieve the right forms. 
 Apart from the $I_n$ fibers, we find a perfect match between the centers of the covering algebra and their associated congruence subgroup. In the $su(n)$ case we also find compatibility to $\Gamma(n)$ groups, that one expects to encode $\mathbb{Z}_n\times \mathbb{Z}_n$ centers. In more in general, there might be the possibility, for some torsion to only mod out a subcenter of some $su(n)$ algebra, when some $\mathbb{Z}_{n \cdot m}$ factor is present. The embedding of the torsion inside the $su(n)$ center is hardly visible from the monodromy picture but in the resolved geometry, which we will comment on in Section~\ref{ssec:hightorsion}. 
 \begin{table}[t!]
 \begin{center}
 \begin{tabular}{|c|c|c|c|c|c|}\hline
   Fiber &$(f,g,\Delta)$ & Monodromy & Subgroups & algebra & center \\ \hline
     $II^*$ & $(\geq 1, 1,2)$ & $ \left( \begin{array}{cc}1 & 1 \\ -1 & 0 \end{array}\right)	$ & -   &$ -$ & - \\ \hline
 $I_n$ & $(0,0,n)$ & $ \left( \begin{array}{cc} 1 & n \\ 0 & 1 \end{array}\right)	$ & $\Gamma_1(n), \Gamma(n)$   &$ su(n)$ & $\mathbb{Z}_n$ \\ \hline
  $III$ & $(\geq 1, \geq 2,3)$ & $ \left( \begin{array}{cc} 0 & -1 \\ 1 & 0 \end{array}\right)	$ & $\Gamma_1(2)$    &$ su(2)$ & $\mathbb{Z}_2$ \\ \hline
   $IV$ & $(\geq 2,2,4)$ & $ \left( \begin{array}{cc} 0 &  1 \\ -1 & -1 \end{array}\right)	$ & $\Gamma_1(3)$    &$ su(3)$ & $\mathbb{Z}_3$ \\ \hline
 
    $I_{2n}^*$ & $(\geq 2,3,6+2 n)$ & $ \left( \begin{array}{cc} -1 &  -2n \\ 0 & -1 \end{array}\right)	$ & $\Gamma_1(2),\Gamma(2) $    &$ so(8+4n) $ & $\mathbb{Z}_2\times \mathbb{Z}_2$ \\ \hline
 
   $I_{2n+1}^*$ & $(\geq 2,3,7+2 n) $ & $ \left( \begin{array}{cc} -1 &  -2n-1 \\ 0 & -1 \end{array}\right)	$ & $\Gamma_1(2),\Gamma_1(4) $    &$ so(10+4n)$ & $\mathbb{Z}_4$ \\ \hline
 
   $IV^*$ & $(\geq 3,  4,8)$ & $ \left( \begin{array}{cc} -1 & -1 \\ 1 & 0 \end{array}\right)	$ & $\Gamma_1(3)$    &$ E_6$ & $\mathbb{Z}_3$ \\ \hline
 
    $III^*$ & $( 3, \geq 5,9)$ & $ \left( \begin{array}{cc}0 & -1 \\ 1 & 0 \end{array}\right)	$ & $\Gamma_1(2)$    &$ E_7$ & $\mathbb{Z}_2$ \\ \hline
 
     $II^*$ & $( 4, \geq 5,10)$ & $ \left( \begin{array}{cc}0 &-1 \\ 1 & 1 \end{array}\right)	$ & -   &$ E_8$ & - \\ \hline 
 \end{tabular}
 \caption{\label{tab:fibermono}{\it Summary of ADE fiber types, their Weierstrass singularity and associated $sl(2,\mathbb{Z})$ monodromy. Column four shows compatible $\Gamma_1(n)$ and $\Gamma(n)$ subgroups matching the center  of the associated algebra. $I_n$ monodromies are both compatible with $\mathbb{Z}_n$  and $\mathbb{Z}_n \times \mathbb{Z}_n$ torsion points. }}
 \end{center}
 \end{table} 
When considering larger torsion, the options of compatible ADE fibers  get more sparse but are still unbounded in principle, thanks to the $su(n)$ factors.  In the next section, we want to switch gears and investigate  whether 6 dimensional anomalies can constrain the order $n$ in a meaningful way.
\subsection{6D SUGRA with Large Putative Torsion} 
From a field theory perspective, there is  no direct reason why torsion of higher orders should not exist, given that there is always some $su(n)$ covering algebra with the given center that can satisfy the strong 6d SUGRA anomalies. These constraints are given as:   
\begin{align} 
\label{eq:6dAnomalies} 
\begin{array}{lcl}
  H-V+29T=273 \, , \qquad  9-T=a  \cdotp a \, , \qquad& \qquad &   (\text{pure-gravitational}) \\
  \textstyle{ -\frac{1}{6}}\left( A_{adj_\kappa}-\sum_{\mathbf{R}} x_{\mathbf{R}} A_{\mathbf{ R}}\right)=a \cdotp    b_{\kappa} \, , &  &  (\text{Non-Abelian-gravitational}) \\ 
  B_{adj_\kappa} - \sum_{\mathbf{R}} x_{\mathbf{R}} B_{\mathbf{R}} = 0\,,& &  (\text{Pure non-Abelian}) \\
  \textstyle{\frac{1}{3}} \left( \sum_{\mathbf{R}} x_{\mathbf{R}} C_{\mathbf{R}}-  C_{adj_\kappa}  \right) =  b_\kappa^2 \, , &&
\end{array}
\end{align}
with $H,V$ and $T$ being the number of massless hyper-, vector-and tensor multiplets in the theory and $x_\mathbf{R}$ the number of hypermultiplets in representations $\mathbf{R}$. The anomaly coefficients $a$ and $b_\kappa$ transform as $SO(1,T)$ vectors and the anomalies can be evaluated using the following group theory coefficients:\footnote{We do not consider su(2) and su(3) groups that have different $C$ and vanishing $B$ and $E$ type coefficients.}
\begin{align}
\text{
\begin{tabular}{|c|c|c|c|c|c|} \hline
 Representation & Dimension & $A_{\mathbf{R}}$ & $B_{\mathbf{R}}$ & $C_{\mathbf{R}}$ & $E_{\mathbf{R}}$  \\ \hline
 Fundamental & $n$ &1 & 1& 0& 1 \rule{0pt}{1Em}\\ \hline
 Adjoint & $n^2-1$ &$2n$ & $2n$  & 6 & 0 
 \\ \hline 
\end{tabular}
} \, .
 \end{align}
 To match the F-theory geometry we make the following identification of base divisors $a\sim K_b$ and $b_\kappa \sim \mathcal{Z}_\kappa $ of $su(n)_\kappa$ whereas $\cdot$ denotes the intersection in the base.\\
When considering $su(n)/\mathbb{Z}_n$ groups, there are several massless representations  that can be exclude straight away due to their incompatible $\mathbb{Z}_n$ center charges. These  include fundamentals and (single times) antisymmetric representations as those are associated to $su(n) \rightarrow su(n+1)$ and $su(n) \rightarrow so(2n)$ algebra enhancements at codimension two  that have incompatible centers for large enough $n$\footnote{The later ones have have at most an $\mathbb{Z}_2$, $\mathbb{Z}_2 \times \mathbb{Z}_2$ or $\mathbb{Z}_4$}. Lets exclude anti-symmetric matter directly and reformulate the anomalies as the following conditions on fundamental and adjoint hypermultiplets:
\begin{align}
x_{fund} = 2 n (x_{adj}-1) \, , \quad x_{adj} = g = 1+\frac12 (\mathcal{Z}^2-\mathcal{Z}\cdot K_b^{-1}) \, ,
\end{align}
where $g$ is the genus of the divisor $\mathcal{Z}$ in the base.  Lets consider solutions, compatible with a putative $su(n)/\mathbb{Z}_n$ group over the divisor $\mathcal{Z}$. A first solution is that of  theory, with only adjoint hyper multiplets, realized by a genus one curve of self-intersection zero. This leaves only the pure gravitational anomaly  that can be solved, given the following amount of pure neutral hypermultiplets:
\begin{align}
H_{neutral}= 272 + n - 29 T \, .
\end{align}
 Lets try to engineer that model more concretely in F-theory. As a base we choose $dP_9$ and the $su(n)$ divisor to be $z=0: \mathcal{Z} \in K_b^{-1}$ to guarantee it to be of genus $g=1$. The factorization of a Tate model
\begin{align}
y^2+x^3+a_1 yx + a_2 x^2+a_3 y + a_4 x + a_6 \, ,\qquad [a_i] \in K_b^{-i} \, ,
\end{align}
is given as:
\begin{align}
a_2 \rightarrow b_1 z \, \quad a_3 \rightarrow c_0 z^3\, , \quad a_4 \rightarrow d_0 z^4\, ,\quad a_6 \rightarrow e_0 z^6 \, ,
\end{align}
with the $c_0, d_0, e_0$ being generic non-vanishing constants. The above factorization, gives an $su(6)$ over $z=0$, without any further codimension two loci since $(K_b^{-1})^2=0$.  By further setting $e_0$ and $c_0$ to zero, this singularity can be enhanced to $su(7)$ and $su(8)$ respectively or by  further setting $b_1=0$  increasing the singularity to $su(9)$. This spectrum admits only a single adjoint representation in its massless spectrum and looks naively like a possible $su(n)/\mathbb{Z}_n$ model for values $n>4$. Especially the later one admits a torsion point of order three as it is a variant of the Schoen \cite{persson,Bouchard:2007mf} but not of order nine as the massless spectrum might suggest.
\\\\
A second solution to the anomaly constraints above is to put the $su(n)$ factor over a $\mathbb{P}^1$ of self-intersection -2 in the base, hosting no adjoint hypermultiplets but instead 2n-fundamentals. In order to get that spectrum consistent with the enhancement rules we introduce another $su(n)$ gauge factor, over another -2 curve giving rise to the enhancement:
\begin{align}
su(n)_1 \cap su(n)_2 \rightarrow su(2n) \, ,
\end{align}
consistent with the $\mathbb{Z}_n$ factor. This configuration can be continued with up to k $su(n)$ factors, intersecting each other as:
\begin{align*}
-su(n)_1-su(n)_2-\ldots su(n)_k- \, ,
\end{align*}
forming the structure of an affine su(k) Dynkin diagram in the base. The resulting gauge group, is of $su(n)^k$ type with k, bi-fundamental representations of type $(\mathbf{1},\mathbf{N}_i, \overline{\mathbf{N}}_{i+1},\ldots,\mathbf{1})$ consistent with an naive $\mathbb{Z}_n$ factor. 
Such a configuration can cancel the gravitational anomaly given the existence of
\begin{align}
H_{neutral} = 273 - k - 29 T \, ,
\end{align}
neutral hyper and tensor multiplets present. 
\\\\
All the above models have a massless spectrum which is seemingly consistent with an $su(n)/\mathbb{Z}_n$ gauge group or multiple copies of those. It seems here that the pure gauge theory anomalies do not give much guidance, on what bounds on the the fundamental group we might expect. However we note, that we mainly argued with the massless sub-sector of the theory and there is a priori no reason to assume, that massive modes must respect the putative $su(n)/\mathbb{Z}_n$ group as well. It is arguably more convincing, that the massive modes respect the (co-)charge refinement that is explicitly enforced by the torsion Shioda map and the reduced monodromy of $\Gamma_{1}(n)$, that also acts non-trivial on (p,q)-strings as argued in Section~\ref{ssec:Fintro}.

\subsection{Higher order and Non-Prime Torsion}
\label{ssec:hightorsion}
In this section we want to comment  on torsion points and their field theory counterparts in the case, when the torsion affects only a sub-center of the gauge algebra as well when it is non-prime and how this is realized in the geometry itself.\\\\ 
Lets consider e.g. the case when only a $\mathbb{Z}_m$ sub-factor of some $su(n \cdot m)$ algebra becomes gauged. This is visible e.g. in the $\mathbb{Z}_2$ torsion model by further specializing $a_4 \rightarrow b_2^2$, which enhances the $su(2)$ to an $su(4)$. The spectrum however is still constraint to an $su(4)/\mathbb{Z}_2\sim SO(6)$ sub-factor, consistent with the existence of antisymmetric $\mathbf{6}$-plet states . This is also consistent with the $\mathbb{Z}_2$ symmetry among 0-and $\mathbb{Z}_2$ torsion which dictates the torsion section to intersect the second $su(4)$ resolution divisor, leaving a torsion Shioda map of the form:
\begin{align}
\Sigma(S_1)=S_1-Z+ \frac12 (f_1 + 2 f_2 + f_3 ) \, .
\end{align}
The factor $1/2$ shows, that the torsion Shioda map is indeed of order two, which gauges the $\mathbb{Z}_2$ sub-center of the $su(4)$ algebra. This does not allow for $\mathbf{4}$-plets with weight $(1,0,0)$  but the anti-symmetric 6-plets with $(0,1,0)$ weight, which can be seen in the associated Weierstrass model explicitly \cite{Mayrhofer:2014opa}. It is easy to extend this observation by noting, that globally, there is the freedom to redefine the zero-and order k-torsion section. This symmetry on the other hand, must extend to the appearing gauge algebra factors and hence a cyclic automorphism of the affine Dynkin diagram itself \cite{Grimm:2015wda}. Indeed this can be geometrically seen by noting that each order $k-$ section can intersect an $su(n\cdot k)$ at the k-th-node. The inverse Cartan  matrix of the $su(n \cdot k)$ then comes with a factor $\frac{1}{det(G)}=\frac{1}{n k}$. However, the $k$-th. row of $C^{-1}$ comes with a k-multiple and hence, the overall  contribution of the $su(n k)$ is only $n-$ fractional. Hence the torsion Shioda map $\Sigma(S_k)$, effects only a sub-center.\\\\ 
Starting from the torsion forms, it is interesting to observe that, with order $MW(Y)_{tor}=\mathbb{Z}_n$ for $n>3$, several additional singularities beyond the expected minimal $su(n)$ start appearing. To consider this effect in more detail, we go back to the $\mathbb{Z}_4$ example with general Weierstrass model:
\begin{align}
\begin{split}
&y^2 + b_1 x y + b_1 b_2 y =  x^3 + b_2 x^2  \,, \qquad [b_1] \in K_b^{-1}\, ,  [b_2] \in K_b^{-2} \, , \\
&f_4 =  -\frac{1}{48} b_1^4 + \frac13 b_1^2 b_2 - \frac13 b_2^2,  \, , \quad  g_4 = - \frac{1}{864} (b_1^2 - 8 b_2) (b_1^4 - 16 b_1^2 b_2 - 8 b_2^2) \, , \\ 
& \Delta=-\frac{1}{16 }b_1^2  b_2^4 (b_1^2 - 16 b_2) \, . 
\end{split}
\end{align}
The above geometry suggests a minimal $(su(2)\times su(4))/\mathbb{Z}_4$ gauge group. The presence of the $su(2)$ factor is especially surprising, as it does not appear in the $\Gamma_{1}(4)$ modular curve, which we discussed in Section~\ref{ssec:z4}. From the argument above, it seems also puzzling how the torsion section can consistently intersect the local $su(2)$, although being of order $4$. Therefore  we will analyze the resolved model in more detail in the following, based on the codimension two resolution with PALP id  $(3145,0)$\cite{Braun:2014qka,Oehlmann:2016wsb
} and equation:
\begin{align}
p_1=& f_{2,1} f_{1,1} f_{2,2} f_{2,3}^2 z_0^2 + b_1 f_{2,1} f_{2,2} f_{2,3} z_1 z_3 + b_2 f_{1,1} z_2^2 \, , \\
p_2=& f_{2,2} z_1^2 + f_{2,1} z_3^2 + f_{1,1} z_0 z_2 \, ,
\end{align}
with $z_i$ toric sections and $f_i$ some resolution divisors that have Stanley-Reisner ideal:
\begin{align}
SRI: \{&  f_{2,2} z_{2}, z_{0} z_{2}, z_{1} z_{2}, f_{2,1} z_{2}, f_{2,3} z_{2}, z_{3} z_{2}, z_{0} z_{1}, f_{1,1} z_{1}, f_{2,3} z_{1}, 
 z_{1} z_{3}, \nonumber \\ & f_{2,2} z_{3}, z_{0} z_{3}, f_{1,1} z_{3}, f_{2,3} z_{3}, f_{2,2} z_{0}, f_{1,1} f_{2,2}, f_{1,1} f_{2,3}, f_{2,1} z_{0},
  f_{2,1} f_{1,1} \} \, .
\end{align}

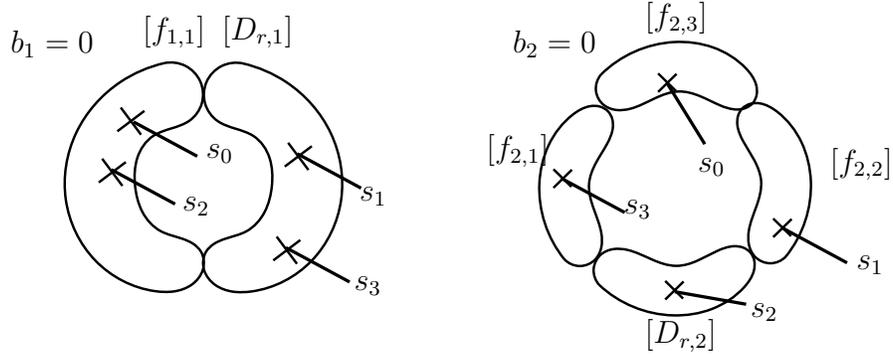
\begin{figure}
\begin{center}
\begin{picture}(0,150)
\put(-150,20){\includegraphics[scale=2.3]{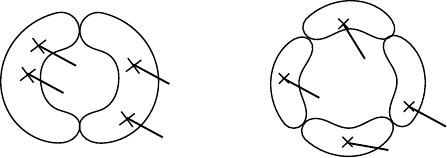}}
\put(-120,125){$[f_{1,1}]$}
\put(-90,125){$[D_{r,1}]$}
\put(-170,120){$b_1=0$}
\put(-96,80){$s_0$}
\put(-105,60){$s_2$}
\put(-38,64){$s_1$}
\put(-40,30){$s_3$}

\put(20,120){$b_2=0$}
\put(10,80){$[f_{2,1}]$}
\put(62,59){$ s_3 $}
\put(70,130){$[f_{2,3}]$}
\put(90,75){$ s_0 $}
\put(140,75){$[f_{2,2}]$}
\put(150,38){$s_1$}
\put(70,10){$[D_{r,2}]$}
\put(110,20){$s_2$}
\end{picture}
\caption{{\it \label{fig:Z4fiber}
Depiction of the generic resolved $\mathbb{Z}_4$ torsion model and its intersection with the torsion sections $s_i$. The intersection pattern with the resolution divisors is compatible with the cyclic order four movement for the $su(4)$ fibers and an order two subgroup for $su(2)$ fibers.
}}
\end{center}
\end{figure}
The resolved fibers are shown in Figure~\ref{fig:Z4fiber} together with the intersection of the four torsion sections $\{ z_0, z_1, z_2, z_3 \}$ which we call $s_{i}$ denoted by their order $i$ in the MW group law\cite{Grimm:2015wda}. In the $su(4)$ it can be observed, that the section intersect in the same order as the Dynkin diagram whereas in the $su(2)$ case they intersect modulo two. This suggests, the presence of the $su(2)$ algebra, due to the non-trivial $\mathbb{Z}_2$ subgroup of $\mathbb{Z}_4$. This can also be observed via the two different torsion Shioda maps $\Sigma(S_i)$, that give a trivial $i$-torsional divisor in the $NS$ lattice via the expressions:
\begin{align}
\Sigma(S_1) =&  [s_1]-[s_0]-\frac12 f_{1,1} + \frac14(f_{2,1} +2 f_r +3 f_{2,2} ) \, ,\\
\Sigma(S_2) = & [s_2]-[s_0]+  \frac12( f_{1,1}  +2 f_r  + f_{2,2}) \, ,
\end{align}
with $f_r= -(f_{2,2}+f_{2,3}+f_{2,1})$.
We observe the first  Shioda map to give a 4-torsion element whereas the second one is only two-torsion, that only sees the $\mathbb{Z}_2$ subgroup of $su(4)$ but not the $su(2)$ at all. 
The above model can be completed to a 6D SUGRA model, which requires the absence of non-flat fibers, or (4,6,12) points, that are generically present over $b_1 = b_2 =0$ and hence requires $(K_b^{-1})^2=0$ which can be achieved over a rational elliptic surface as the base. In order to make the model simple, we can factor $a_2 \rightarrow c_1 b_1$, giving rise to two copies of $su(4)$ factors discussed above. This model can also be arranged as a variant of the Schoen manifold  \cite{Bouchard:2007mf}, which, in addition to the $(su(4)^2 \times su(2))/\mathbb{Z}_4  $ admits a non-higgsable U(1) factor. The full matter spectrum of the model is specified in Table~\ref{tab:FoldingQuotient} and can be shown to satisfy gauge and gravitational anomalies
\begin{table}[t!]
\begin{center}   
\begin{tabular}{|c|c||c|} \hline
Gauge group: & $  (su(4)^2 \times su(2))/\mathbb{Z}_4 \times U(1) $ & $  (su(6) \times su(3)\times su(2))/\mathbb{Z}_6 $  \\ \hline
$(h^{1,1},h^{2,1})$&$(19,19)$ & $(19,19)$    \\ \hline
$H $: & $ 2\times \mathbf{15}\oplus \mathbf{3}\oplus 20\times \mathbf{1}$ & $   \mathbf{35}\oplus \mathbf{8}\oplus \mathbf{3} \oplus 20\times \mathbf{1}$  \\ \hline
$T : $&$ 9$  &$ 9$  \\   \hline
\end{tabular}  
\end{center}
\caption{
\label{tab:FoldingQuotient} {\it Summary of  massless 6D F-theory spectra with $\mathbb{Z}_4$ and $\mathbb{Z}_6$ fundamental group, geometrically realized by the Schoen manifold.}
}
\end{table} 
The observations above, are pathological for non-prime $\mathbb{Z}_n$ order torsional models. The presence of $su(m)$ gauge factors, where $m$ are divisors of $n$ forces $(4,6,12)$ points over generic Fano bases  that can be avoided over  dP$_9$.. The maximal case possible are $\mathbb{Z}_6$ for a single factor and $\mathbb{Z}_3 \times \mathbb{Z}_3$ for two factors, as concretely constructed in \cite{Aspinwall:1998xj}. The Weierstrass coefficients of the former one is given as
\begin{align} 
\begin{split}
 y^2+a_1 xy &+\frac{1}{32}(a_1 - b_1)(3 a_1+ b_1)(a_1 + b_1) = x^3+\frac18(a_1 - b_1)(a_1+b_1) x^2\\
f=&\frac{1}{192} b_1 ( 3 a_1^3 - 3 a_1^2 b_1-3 a_1 b_1^2 - b_1^3 )   \\
g=&    \frac{1}{110592} (3 a_1^2 - 6 a_1 b_1-b_1^2)(9 a_1^4 -6  a_1^2 b_1^2 -24 a_1^3 b_1-11 b_1^4      ) \\
\Delta=& \frac{1}{2^{24}} (a_1-5 b_1)(3 a_1 + b_1)^2(a_1+b_1)^3(a_1 - b_1)^6  \, , \quad a_1,b_1 \in \mathcal{O}(K_b^{-1})\, , 
\end{split}
\end{align}
and is also naturally described by a Schoen manifold, in order to keep all (4,6,12) points absent. Note that the generic $I_6, I_3$ and $I_2$ singular fibers expected from the cusp points of the modular curve, as discussed in Section~\ref{sec:mcurves} and depicted in Figure~\ref{fig:X16} are present. This model gives rise to a good F-theory vacuum in six dimension with $(su(6)\times su(3) \times su(2))/\mathbb{Z}_6$ gauge group, whose massless spectrum is summarized in Table~\ref{tab:FoldingQuotient}. \\\\
The modular curve perspective gives a good understanding of which minimal singularities are to be expected in certain torsion models. Lets go to a higher $n$-torsion point say $\mathbb{Z}_8$ e.g. over a $\mathbb{F}_0\sim \mathbb{P}^1_{[t_0,t_1]}\times \mathbb{P}^1_{[s_0,s_1]}$ base which we derive explicitly in Appendix~\ref{app:ellipticsurfaces}  and analyze the associated Weierstrass model directly. We repeat the model here, as:
\begin{align}
\label{eq:z8example}
 &y^2 + (s_0 s_1 t_0 t_1 + (s_0 t_0 - s_1 t_1) (2 s_1 t_1-s_0t_0)) xy + s_0 s_1^3 t_0 t_1^3 (s_0 t_0 - 2 s_1 t_1) (s_1 t_1 - s_0 t_0) y \nonumber
  \\ &= 
 x^3 + s_1^2 t_1^2 (s_0 t_0 - 2 s_1 t_1) (s_1 t_1 - s_0 t_0)x^2 \, ,
\end{align}
with discriminant:
\begin{align}
\Delta =  \frac{-1}{16}s_0^2 s_1^8 t_0^2 t_1^8 (s_0 t_0 - 2 s_1 t_1)^4 (s_0 t_0 - 
   s_1 t_1)^8 (s_0^2 t_0^2 - 8 s_0 s_1 t_0 t_1 + 8 s_1^2 t_1^2) \, .
\end{align}
This model admits two  $I_2$, one $I_4$ and three $I_8$ fibers  
 as well as two (8,12,24) codimension two singularities. Therefore, the $(su(8)^3 \times su(4) \times su(2)^2)/\mathbb{Z}_8$ six dimensional SUGRA model can not exist. In the next section we show, however that a related model in eight dimensions can exist.
\subsection{K3: 8-dimensional Exceptions}
\label{ssec:K3}
The only way to avoid the (8,12) singularities at codimension two is by restricting the base to be one-dimensional, that is a K3 surface.    
This allows some additional possible higher order torsion models to exist exclusively as 2-folds, as opposed to 3- and 4-folds. Hence their underlying 8D SUGRA theories, obtained from F-theory comprise of isolated theories as we discuss in the following. The $\mathbb{Z}_8$ model is one of the few exceptional torsion models, only\footnote{As opposed to Calabi-Yau fibrations of higher dimension.} present in elliptic K3's. The groups and their fiber configurations are summarized in Table~\ref{tab:MWK3}, that is adapted from a table in \cite{MP}.
\begin{table}[h!]
\begin{center}
\begin{tabular}{|c|c|}\hline
MW$_{tor}$ & Fibers \\ \hline
$\mathbb{Z}_7$ &$3 I_7$ \\ \hline
$\mathbb{Z}_8$ &$2 I_8+I_4+I_2$ \\ \hline
$\mathbb{Z}_6 \times \mathbb{Z}_2$ &$  3  I_6+3 I_2$ \\ \hline
$\mathbb{Z}_4 \times \mathbb{Z}_4 $& $ 6 I_4 $  \\ \hline  
\end{tabular}
\caption{{\it \label{tab:MWK3} Summary of large torsion models, that can not be realized in 3- and 4-folds and their fibers.  All of them are extremal K3's.}}
\end{center}
\end{table}
As can be found, all elliptic K3's are of extremal type which means that their N$\acute{\text{e}}$ron-Severi group has maximal possible rank 20. In short, no elements of $h^{1,1}(Y)$ can contribute to complex structure deformations which results in a rigid geometry. 
  In fact, the  $T=\mathbb{Z}_8$ threefold given in \eqref{eq:z8example} can be viewed as an enforced K3-fibration over a base $\mathbb{P}^1$ in terms of $[t_0,t_1]$ coordinates that preserves the torsion. Deleting the $\mathbb{P}^1$ by effectively setting the coordinates to one, gives the consistent K3 with the correct fibers.\\ 
In this regard, also their associated 8D F-theory vacua are special. Their rigidity, does not allow for a minimal SUSY preserving compactification to lower dimension\footnote{Note that trivial, fibrations over a torus or K3 \cite{Kimura:2016gxw} are still possible.} compatible with the torsion.  Moreover, the rigidity, also does not allow for a stable degeneration limit at finite distance in the moduli space into two rational elliptic surfaces, as those require at least one free complex structure modulus to assign the proper scaling. Therefore, these models do not have a (geometric)\footnote{In \cite{Mayrhofer:2017nwp} non-geometric heterotic-F-theory dual models in eight and lower dimensions have been proposed, some of them also admitting non-crepant singularities.
} heterotic dual either. This is consistent with the fact, that the groups do not fit into $E_8\times E_8$, which is also reflected by the fact, that none of the torsion factors, are discrete finite subgroups of $E_8$. 
\\
\begin{figure} 
\centering
 \includegraphics[width=6in]{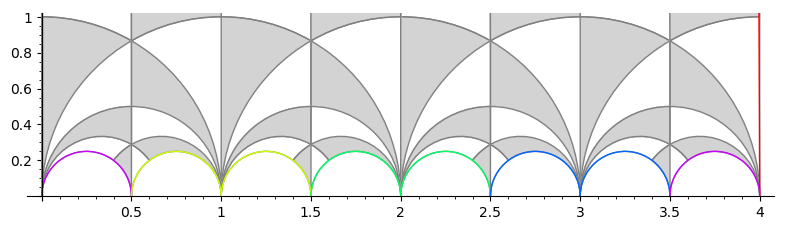} \\
 \includegraphics[width=3in]{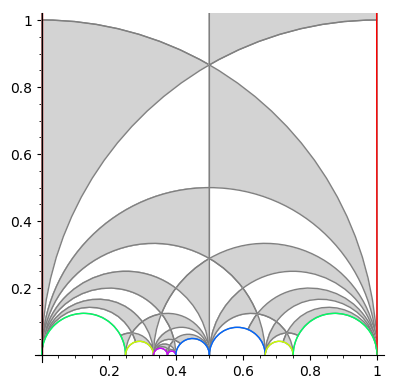} 
  \includegraphics[width=3in]{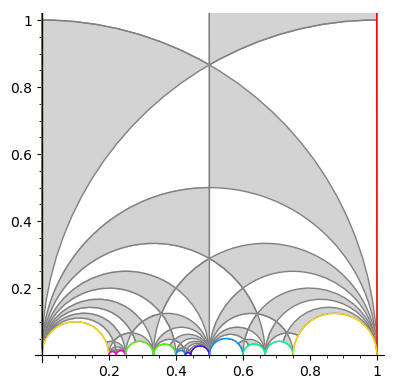} 
 \caption{\label{fig:Z9} {\it Triangulations of $X(4)$, $X_1(8)$ and $X_1(9)$. Colors highlight identified lines, allowing to read off the cusp points and their widths.}}
\end{figure}
The same conclusions can be obtained, by simply considering the modular curves of the higher torsion models which naturally explains the absence of even larger torsion in K3.  In Figure~\ref{fig:Z9} we depicted the three modular curves that correspond to $\mathbb{Z}_4 \times \mathbb{Z}_4$, $\mathbb{Z}_8$ and $\mathbb{Z}_9$ torsion model explicitly. The $\mathbb{Z}_4 \times \mathbb{Z}_4$ modular curve, admits six cusp points at
\begin{align}
S^{(4\times4)}_{\text{cusp}}: \{ 0^{(4)}, 1/2^{(4)}, 1^{(4)}, 2^{(4)}, 3^{(4)}, \infty^{(4)} \} \, ,
\end{align}
where we highlighted their widths as superscripts. The associated fibers are consistent with the expectation from the K3 classification. Similarly, the $\mathbb{Z}_8$ curve, discussed earlier admits six cusps 
\begin{align}
S^{(8)}_{\text{cusp}}: \{ 0^{(8)}, 1/4^{(2)}, 1/3^{(8)}, 3/8^{(1)}, 1/2^{(4)}, \infty^{(1)} \} \, ,
\end{align}
consistent with K3 construction as well. Finally we depicted also the $\mathbb{Z}_9$ case in Figure~\ref{fig:Z9}. There are eight cusps in total at
\begin{align}
S^{(9)}_{\text{cusp}}: \{ 0^{(9)}, 2/9^{(1)}, 1/4^{(9)}, 1/3^{(3)}, 4/9^{(1)}, 1/2^{(9)}, 2/3^{(3)}, \infty^{(1)} \} \, .
\end{align} 
Collecting all singular fibers, we end up with three $I_9$ and two $I_3$ fibers in total. Resolving these singularities while adding the class of the generic fiber and base required $30$ independent K{\"a}hler parameters. This clearly over shots the maximum of 20 that  K3  has at its disposal by large. With view on F-theory, this also shows explicitly why there can be no eight dimensional SUGRA theories, with a non-simply connected gauge group larger eight, or $\mathbb{Z}_4 \times \mathbb{Z}_4$. 

\subsection{Bounds on Calabi-Yau Quotient Torsors}

These results also reveal constraints on a certain class of non-simply connected, genus-one fibered Calabi-Yau 3-folds 
$\widehat{Y}_3$ which are relevant for heterotic \cite{Donagi:1999ez,Donagi:2000zf,Braun:2017feb} and F-theory compactifications \cite{Anderson:2018heq,Anderson:2019kmx}.
The construction of the 3-folds is described in detail in \cite{Donagi:1999ez,Bouchard:2007mf}.
We start with a smooth elliptic 3-fold $X_3 \rightarrow B_2$,
and assume that we have the following extra structure:
\begin{itemize}

\item{An automorphism $\al$ of $B_2$ which is of finite order, and which is compatible with the fibration structure.}
\item{A non-identity element $P\in MW(Y/B)$ satisfying:
\begin{itemize}

\item{$P$ is fixed by $\si$, and $nP = 0$ in MW.}
\item{$P$ is not fixed by $\si$, and the sum of the elements in the $\al$ orbit of $P$ is 0.}

\end{itemize}}

\end{itemize}

We then combine the automorphism of the base with a translation by $P$ along the fibers to obtain a finite order, fixed point-free automorphism of $Y_3$, that we denote $\tilde{\al_P}$.
We can take a simultaneous quotient of $Y_3$ and $B_2$ by the action of $\ip{\tilde{\al_P}}$ and $\ip{\al}$, respectively, we obtain a new fibration $Y_3/\ip{\tilde{\al_P}}\rightarrow B_2/\ip{\al}$.

The resulting quotient geometries have two central properties that we want to comment on.
\begin{enumerate}

\item Although $Y_3//\ip{\tilde{\al_P}}$ is itself smooth, $B_2/\ip{\al}$ has isolated singularities over the fixed points of $\al$ action.
The fibration has multiple fibers\footnote{In the context of F-theory, these fibers also appeared in the duality to the CHL string \cite{  deBoer:2001wca} and classification of little string theories \cite{Bouchard:2007mf}.} over these singular points  \cite{Griffiths} .

\item Second, the new fibration no longer has a global section. 
This is immediate from the fact that we have multiple fibers over the singularities in the base.
The zero section of the original fibration becomes a multisection of degree $n$ in the quotient fibration.

\end{enumerate}
In the physics of F-theory, the $n$-multisections have been shown to lead to discrete $\mathbb{Z}_n$ gauge symmetries  in the 6D SUGRA theory. The fixed points on the other hand carries  superconformal matter theories coupled to gravity. In \cite{Anderson:2018heq,Anderson:2019kmx} consistency of such theories for all $\mathbb{Z}_n$ quotients have been shown. \\\\
The bounds on the MW torsion groups therefore also constrains the possibilities for this construction using torsion points.  In the physics of F-theory, these constraints can be translated into bounds on discrete symmetries and gaugings of superconformal matter.
In particular the large order quotients, $n>4$ turn out to be all quotients of the Schoen manifolds, classified in \cite{Bouchard:2007mf} whose F-theory physics has been investigated in \cite{Anderson:2019kmx}.Therefore, the orders of the quotients can not exceed the symmetries possible in rational elliptic surface and hence these are all possible quotient manifolds\footnote{Alternatives might be quotients of type $(K3 \times T^k)/\mathbb{Z}_n$ \cite{deBoer:2001wca,Cota:2019cjx} as well as quotients, that do not require a finite Mordell-Weil group.} possible via that construction.

Furthermore, the technical results we use to bound MW torsion can also be used to bound the order of fibration-preserving automorphisms of a Calabi-Yau 3-fold.
Let $\pi : Y \rightarrow B$ be an elliptic 3-fold given by a Weierstrass equation.
Let $L = K(B)$ be the function field of the base and $K = \CC(f,g)$ the field generated by the Weierstrass coefficients.
Then one of the following must be true:
\begin{itemize}
\item{$K = \CC$. In this case, the fibers in the elliptic curve do not vary, since $f,g$ are constants.}
\item{$L/K$ is a finite field extension. In this case, elementary Galois theory tells us the group of automorphisms of $L$ which fix $f,g$ is contained in $S_n$, where $n = [L:K]$.
In particular, there are only finitely many automorphisms of the base which fix $f,g$.}
\item{Otherwise, $L$ has transcendence degree 1 over $\CC$.
In this case, $L$ is the field of meromorphic functions of some Riemann surface $C$, and the inclusion of fields $K \rightarrow L$ means we have a rational map $B \rightarrow C$. Lemma 5.1 below shows that $B$ is flat and $C$ has genus 0, and so $Y\rightarrow B$ is a special fibration.
Resolving indeterminacy of the map $B \rightarrow C$ requires blowing up $B$ in the base points of a pencil of cubics, after which $B$ becomes a rational surface. 
It is not too hard to show that $Y$ is then Schoen, since the fibration $Y \rightarrow B$ factors through the fiber product $Y\rightarrow \tilde{B}\times_ C Y_C$, and minimality of $Y \rightarrow B$ forces that map to be an isomorphism.
Since this construction has been fully analyzed on Schoen manifolds, this shows that any fixed Calabi-Yau 3-fold admits only finitely many quotient torsors. 
}
\end{itemize}

\section{Technical Results}
\label{sec:5}

Our main result is the bound on torsion groups in smooth Calabi-Yau $n$-folds, $n \geq 3$.
The key idea in the proof is the fact that the diagram (2.4) forces $(f,g,\Delta)$ to vanish to order $(4d,6d,12d)$ over points in the indeterminacy locus of $\phi$ which obstruct the existence of a flat crepant resolution of the total space.
To ease exposition, we will prove the theorem for 3-folds specifically and comment on its validity for four-or five folds at the end.

To avoid having to constantly refer to the diagram,
we introduce the phrase ``special fibration".
An elliptic fibration $Y_n \rightarrow B_{n-1}$ is {\bf special} if it fits into a commutative diagram:
\begin{equation}
\label{eq:commut}
\begin{tikzcd}
Y_n \arrow[r, dashed, "\Phi"] \arrow{d}{\pi}
&S_2 \arrow{d}{p}\\
B_{n-1} \arrow[r,dashed,  "\phi"]  &C_1 \, ,
\end{tikzcd}
\end{equation}
satisfying the following conditions:

\begin{itemize}
\item{$S_2 \rightarrow C_1$ is a smooth, minimal elliptic surface.}
\item{The horizontal maps are non-constant, rational maps.}
\item{The vertical maps are proper.}
\item{$\phi$ is flat.}
\end{itemize}

\subsection{Global Lemmas}

{\lem{Let $B$ be a rational variety and $\phi : B \rightarrow C$ be a non-constant rational map.
Then $C$ has genus 0 and $\phi$ is flat.}}

\begin{proof}
Since $\phi$ is non-constant, we can find points $b_1, b_2 \in B$ such that $\phi(b_1) \neq \phi(b_2)$.
Since $B$ is rationally connected, there is a regular map $\PP^1 \rightarrow B$ 
taking $0$ to $b_1$ and $\infty$ to $b_2$.
The composition $\PP^1 \rightarrow C$ is a non-constant rational map,
so it extends to a surjective morphism.
This is only possible if $C$ has genus 0.

To prove flatness of $B \rightarrow \PP^1$, it suffices to prove the map is flat over each point in $\PP^1$.
Any proper open neighborhood of $\PP^1$ has the form $\spec R_0$ for $R_0$ a principal integer domain (PID),
so we can determine whether the map is flat by studying the morphism of algebras $R_0 \rightarrow K(B)$.
Since $R_0$ is a PID, flatness is equivalent to $K(B)$ being torsion-free,
which follows immediately from the fact that the map $B\rightarrow \PP^1$ is non-constant and $K(B)$ is purely transcendental.

\end{proof}

The fundamental line bundle of an elliptic fibration $\pi : Y \rightarrow B$ is $R^1\pi_* \Oo_Y$.
We denote it by $\Ll_{Y/B}$.
We write $\om_Y$ (resp. $\om_B$) to denote the canonical bundle of $Y$ (resp. $B$).

{\lem{Let $B$ be a rational surface and $\pi : Y \rightarrow B$ a special elliptic 3-fold.
Then $\Ll_{Y/B} \cong \phi^*(\Ll_{S/\PP^1})$.}}

\begin{proof}

The conditions in the definition of a special 3-fold allow us to use Prop. III.9.3 in \cite{Hartshorne}
to compute:
\begin{align} \Ll_{Y/B} = R^1 \pi_* \Oo_Y = R^1 \pi_* \Phi^*\Oo_S = \phi^*R^1 p_* \Oo_S = \phi^* \Ll_{S/B} \, . \end{align}

\end{proof}

{\lem{Let $\pi : Y \rightarrow B$ be as above and let $d$ be the degree of $\Ll_{S/\PP^1}$.

Then $\om_Y \cong \om_B \otimes \phi^*(\Oo_{\PP^1}(1))^{\otimes d}$.
}}

\begin{proof}
This follows from the canonical bundle formula for elliptic fibrations,
together with the computation from the previous lemma.

\end{proof}

{\prp{Let $\pi : Y \rightarrow B$ be a special elliptic 3-fold.
Assume that $B$ is a minimal rational surface and $\om_Y$ is trivial.
Then $\phi$ is not a morphism.}}

\begin{proof}

There are no non-constant morphisms $\PP^2\rightarrow \PP^1$,
so we may assume $B \cong \FF_n$ for $0 \leq n \leq 12$.
Note that in all cases $NS(B) \cong \ZZ\oplus \ZZ$,
the class of the canonical bundle is $(-2, -n)$,
and $[K_B ]\cdot [K_B] = 8$.

If $\phi\inv(\Oo(1)) = \om_B$, then the number of points in the locus of indeterminacy (counted appropriately) is $[K_B] \cdot [K_B] =8$.
Otherwise, $n$ is necessarily even and $\phi\inv(\Oo(1))^{\otimes 2} \cong \om_B$.
In this case, the number of points in the locus of indeterminacy is $\frac{1}{4} \cdot 8 = 2$. 

In all cases, the requirement $\phi\inv(\Oo(1)) \cong \om_B$ forces the existence of at least one base point.

\end{proof}

\subsection{Local Lemmas}

In this section $k$ is an algebraically closed field,
$R$ a factorial, finitely generated $k$-algebra and $K$ is the fraction field of $R$.

{\df{Let $\mx \subset R$  be a maximal ideal and $I \subset R$ an ideal.
The {\bf order of vanishing of $I$ at $\mx$} is the largest integer $m$ such that $I\subset \mx^m$.}}

{\df{Let $\phi =\frac{p}{q} \in K$,
with $p,q \in R$ relatively prime. 
We can think of $\phi$ as a map $\spec R \rightarrow \PP^1$.
The {\bf locus of indeterminacy} of $\phi$ is $V(p)\cap V(q) \subset \spec R$.}}

{\df{Let $\phi$ be as above, and assume the locus of indeterminacy of $\phi$ contains a closed point $b$.
Let $\mx_b \subset R$ be the corresponding ideal.
We define $m_\phi (b)$ to be the multiplicity of the ideal $(p,q)$ at $\mx_b$.}}

Note that $m_\phi(b)$ makes sense whenever we have a rational map $B\rightarrow \PP^1$ from a normal scheme $B$, since the property is local on $B$.

{\prp{Let $\phi : B \rightarrow \PP^1$ be a non-constant rational map and
$f$ a global section of $\Oo_{\PP^1}(d)$ for some $d>0$.
If $b \in B$ is in the indeterminacy locus of $\phi$,
then $\phi^*(f)$ vanishes to order $m_\phi(b) d$ at $b$.
}} 
\begin{proof} 
The claim is local on $B$, so we assume $B$ is affine,
say $B = \spec R$.
Choosing coordinates $[x_0:x_1]$ on $\PP^1$,
we can express $f(x_0,x_1)$ as a homogeneous polynomial of degree $d$ in $x_0,x_1$.
Furthermore, we can write the map $\phi : B\rightarrow \PP^1$ as $b\mapsto [p(b):q(b)]$, where $p,q \in R$ have no common factors.
In this notation, we have $\phi^*(f)(b) = f(p(b),q(b))$.

Since $f$ is a homogenous polynomial of degree $d$,
$f \in (x_0,x_1)^d \subset k[x_0,x_1]$.
If $b \in B$ is in the locus of indeterminacy of $\phi$,
then $\mx_b^{m_\phi(b)} \supset (p(b),q(b)) = \phi^\#((x_0,x_1))$ so:
\[\phi^*f \in (p(b),q(b))^d \subset \mx_b^{m_\phi(b) d} \] 
\end{proof}

We now easily deduce the following:

{\cor{Let $\pi :Y\rightarrow B$ be a special fibration,
let $d$ be the degree of the fundamental line bundle of $\Ll_{S/\PP^1}$, 
and let $b \in B$ be a point in the indeterminacy locus of $\phi$.
Then the Weierstrass coefficients $f,g$ of $Y$ vanish to order $(4n,6n)$, where $n = dm_\phi(b)$.
}}

\begin{proof}

Commutativity of the square tells us $f_B = \phi^*(f_\PP^1)$ and $g_B = \phi^*(g_\PP^1)$.
The Weierstrass coefficients of the elliptic surface are homogenous polynomials of degree $4d, 6d$ respectively,
proving the claim.
\end{proof}

{\cor{Let $\pi : Y \rightarrow B$ be special elliptic fibration and suppose the order of vanishing of $(f,g)$ does not exceed $(4,6)$
over any point $b \in B$.
Then either $\phi$ is a morphism, or $S$ is rational and the locus of poles and the locus of zeros of $\phi$ intersect transversely.
}}

\begin{proof}

Assume $\phi$ is not a morphism.

An elliptic surface is rational if and only if the fundamental line bundle has degree 1.
If the fundamental line bundle has degree $d>1$,
then the order of vanishing over all points in the locus of indeterminacy is at least $(4d,6d)$.
Thus the condition on $(f,g)$ forces $d =1$, hence rationality of $S$.
If the locus of poles of and zeros meet non-transversely at some point $b$,
then $m_\phi(b)>1$ so $(f,g)$ vanish to order at least $(8,12)$ over $b$.

\end{proof}

\subsection{Summary of Technical Portion}

The results of the previous two sections are summarized in the following theorem:

{\thm{Let $\pi : Y \rightarrow B$ be an elliptic 3-fold satisfying:
\begin{enumerate}
\item{$Y$ has trivial canonical bundle.}
\item{$B$ is a minimal rational surface.}
\item{The order of vanishing of the Weierstrass coefficients $(f,g)$ does not reach
$(8,12)$ over any point $b \in B$.
}
\end{enumerate}

If the fibration is special, then:
\begin{itemize}
\item{$C \cong \PP^1$.}
\item{$S$ is rational, or equivalently $\Ll_{S/C} \cong \Oo(1)$.}
\item{ $\phi^*(\Oo_{\PP^1}(1)) \cong \om_B^\vee$}
\item{ $m_\phi(b) = 1$ for all points in the locus of indeterminacy of $\phi$.} 
\end{itemize}
}}

We make some remarks before proving the main theorem.

\begin{itemize}
\item{The condition on the order of vanishing of $(f,g)$ is precisely the condition needed to guarantee that the Weierstrass model admits a proper, flat crepant resolution, see e.g. \cite{Grassi}.}
\item{Using the birational classification of algebraic surfaces,
we can be more precise: $\phi$ has 8 or 9 points of indeterminacy, with 9 occurring if and only if $B \cong \PP^2$.
If we assume that $B = \PP^2$ e.g.,
it is easy to see that resolving indeterminacy of $\phi$ means blowing up $\PP^2$ at the 9 points in the base locus of a pair of cubics, so the new map $\tilde{\phi} : \tilde{B}\rightarrow \PP^1$ is itself an elliptic fibration  Commutativity of \eqref{eq:commut}  gives us a natural map $\tilde{Y}\rightarrow \tilde{B}\times_{\PP^1} S$.
Minimality of the fibration $\tilde{Y}\rightarrow \tilde{B}$ then forces that map to be an isomorphism, showing that any special $Y\rightarrow B$ is birational to a Schoen manifold.
}

\end{itemize}
\subsection{Application}

{\thm{Let $\pi : Y \rightarrow B$ be an elliptic 3-fold satisfying the hypotheses of the previous theorem.
Then $MW(Y/B)_{tors}$ is isomorphic to one of the following groups:
\begin{align*} \ZZ_n:& \qquad (n = 1,2,3,4,5,6)\, , \\ 
 \ZZ_2 \times \ZZ_{2m}:& \qquad (m = 1,2)\, , \qquad \ZZ_3 \times \ZZ_3 \, . \end{align*} }} 
Note that this list of groups is exactly the list studied in \cite{Aspinwall:1998xj}. 
\begin{proof}

The map $B\rightarrow X_1(n)$ shows that $Y\rightarrow B$ is special.
The degree of $\Ll_{S/C}$, where $C \cong X_1(n)$, can be found in \cite{MP} or computed using the formulae in Appendix~\ref{app:indices}.
To obtain a smooth Calabi-Yau 3-fold,
we need the degree to be exactly 1, which occurs precisely for the torsion groups listed above. 
\end{proof} 
In particular, we note that this argument can be applied to bound Mordell-Weil torsion in Calabi-Yau fibrations of larger dimension:
the map $\phi : B \rightarrow C$ always exists,
and is guaranteed to have nontrivial indeterminacy locus if we assume the fundamental line bundle has positive degree.
The order of vanishing of $(f,g,\Delta)$ over points in the indeterminacy locus is at least $(8,12,24)$ as soon as $\deg \Ll_{S/C} > 1$.
Thus, it is impossible to crepantly resolve singularities of the Weierstrass model and obtain a smooth Calabi-Yau total space. We devote Appendix~\ref{app:ellipticsurfaces} to the explicit construction Weierstrass models beyond the list above. 

\section{Summary and Conclusions}
\label{sec:6} 
In this work we proved that the Mordell-Weil torsion group $T$ of elliptically fibered Calabi-Yau $n$-folds, with $n>2$ can not exceed $T=\mathbb{Z}_6$ and $T=\mathbb{Z}_3 \times \mathbb{Z}_3$. 
We showed explicitly that generic Weierstrass models with higher prescribed torsion have singularities over a codimension two locus in the base which do not admit a crepant resolution. This shows that the list of Weierstrass models that appears in \cite{ Aspinwall:1998xj} contains every possible torsion group that is allowed on a smooth Calabi-Yau. Furthermore we use those results to bound possible quotient Calabi-Yau torsors as well. 
  We use modular curves to interpret those minimal singularities as a direct feature of the congruence groups $\Gamma_1(n)$ and $\Gamma(n)$ of $\mathrm{SL}(2,\mathbb{Z})$. There, the amount and type of minimal singularities can directly be read off from cusps in the fundamental domain of the modular curve of the respective congruence subgroup. We find the order of torsion groups bound by those of rational elliptic surfaces.
On the other hand, K3 manifolds can avoid (8,12) non-crepant codimension two singularities and allow for more possibilities. By connecting the index of the congruence subgroups, with the degree of the fundamental line bundle, we can show however  that also K3 can not exceed $\mathbb{Z}_8$ and $\mathbb{Z}_4 \times \mathbb{Z}_4$ as classified in \cite{MP}.  These  K3  manifolds however are all of extremal type and do not admit any free complex structure deformation.  Therefore they neither have a stable degeneration limit into rational elliptic surfaces, nor can they be used as building blocks for K3 fibered three- or four-folds while preserving the torsion. 
We further interpret these bounds within the physics of F-theory, where torsion realizes  the fundamental group of the global gauge group as a {\it swampland} constraint. We argue that, from pure field theory considerations, the above constraints are surprising as it is possible to construct rather large rank $su(n)$ gauge theories where the massless spectrum respects a putative $\mathbb{Z}_n$ quotient factor.  This might point towards the possibility, that it is the massive spectrum that does not respect the putative first homotopy group, which  is geometrically enforced when torsion is present.   \\
 In addition we observe a variety of surprising effects when large torsion is present that have no field theory explanation yet. The necessity for multiple gauge factors seems especially puzzling from a pure field theory point of view. A possible explanation of this effect might be to view the torsion as a kind orbifold in   $\mathrm{SL}(2,\mathbb{Z})$ of the F-theory torus to an $\Gamma_1(n)$ subgroup requiring the presence of additional gauge theory sectors for consistency of the modularity. 
It would also be interesting to explore whether  Dai-Freed anomalies \cite{Garcia-Etxebarria:2018ajm} 
could be the right framework  to gain a better understanding on why we see these specific gauge algebra factors for a given large fundamental group.
  These bounds however can partially be avoided in isolated 8D SUGRA vacua. These vacua admit at most trivial circle compactifications, but none that preserve only a minimal amount of SUSY and the torsion group. They also do not admit a (geometric) heterotic dual  due to their rigidity and the fact that their minimal gauge group does not fit into $E_8 \times E_8$ (or SO(32)) . It would be interesting, to investigate those eight dimensional exceptions from a field theoretical perspective, maybe in the spirit of \cite{Garcia-Etxebarria:2017crf}. Finally, since all admissible torsion groups must be embeddable into a rational elliptic surface (and into its $E_8$ lattice) it seems plausible, that the heterotic string plays a similar prominent role as in \cite{Lee:2019skh} to explain the constraints we find from the F-theory perspective.

\section*{Acknowledgments}
P. K.O. would like to thank  James Gray, Nikhil Raghuram and Fabian Ruehle for interesting discussions. N.H. would like to thank Steve Trettel for helpful explanations. N.H. and P.K.O. would like to thank Dave Morrison and Markus Dierigl for helpful conversations and suggestions. The work of P. K. O. is supported by an individual DFG grant OE 657/1-1. The research of N. H. was partially supported by the National Science Foundation, grant \# PHY-1620842. P. K. O would like to gratefully acknowledge the hospitality of the Simons Center for Geometry and Physics (and the semester long program, \emph{The Geometry and Physics of Hitchin Systems}) during the completion of this work. The authors would also like to thank the organizers of the 2019 workshop  \emph{The Physics and Mathematics of F-theory} held at Florida State University where this work was initiated. 
\newpage
\appendix

\section{Universal Elliptic Curves with Points of Order $n \leq 8$.}
\label{app:ellipticsurfaces} 
In this appendix we give explicit constructions for the universal elliptic curves with a point of order $n$, for $2 \leq n \leq 8$,
and we show how to extend the natural rational maps $B\rightarrow \PP^1$ over codimension 1 components of the discriminant when $n = 7,8$.
The construction for modular curves is well-known in the arithmetic theory of elliptic curves, see \cite{Mazur78}
See \cite{MP, shioda3} for the generalization to elliptic surfaces.
We are including the explicit constructions since these maps have not been used to study higher dimensional elliptic fibrations and the information we need is in the locus of indeterminacy. 

The ``main" construction applies only to elliptic curves with points of order $n \geq 4$.
For completeness, we also explain what happens when we try to parametrize elliptic curves with a 2 or 3 torsion point.

The construction can be carried out over any ground field - for simplicity, we assume our ground $K$ has characteristic 0 throughout.

\subsection{$n=2,3$}

Let $E/K$ be an elliptic curve and $P \in E(K)$ a 2-torsion point.
We can choose coordinates so:
\begin{align} E : \quad  y^2 = x ( x^2 + ax + b)\, , \qquad P = (0,0)\, , \end{align}
for some $a,b \in K$.
The isomorphism class of the pair $(E, P)$ is determined by $a,b$,
but the choice of $a,b$ are not unique.
Two choices $a,b$ and $a',b'$ give rise to isomorphic pairs if and only if there exists $\lambda \in K^\times$ such that $\lambda^2 a = a'$ and $\lambda^4 b = b'$.
Thus, the modular curve parameterizing pairs $(E,P)$,
where $P$ is a 2-torsion point,
is:
\begin{align}
X_1(2) = \set{(a,b) \in K^2 }/ (a,b) \sim (\lambda^2 a, \lambda^4 b) \, ,
\end{align} 
Similarly, any pair $(E, P)$ with $P$ a point of order 3 is isomorphic to:
\begin{align}
y^2 + axy + by = x^3 \, ,\qquad P = (0,0) \, ,
\end{align}
for some $a,b \in K$.
Two choices $a,b$ and $a',b'$ give rise to isomorphic pairs if and only if there exists $\lambda \in K^\times$ such that $\lambda a = a'$ and $\lambda^3 b = b'$. 
Thus:
\begin{align}
X_1(3) = \set{(a,b) \in K^2 }/ (a,b) \sim (\lambda a, \lambda^3 b) \, .
\end{align}
Both of these quotients are singular - this is due to the presence of finite order points in $\Gamma_1(2), \Gamma_1(3)$.
For $n \geq 4$, there are no more finite order points in the congruence subgroup,
so we obtain smooth modular curves.
As a result, the moduli spaces are better behaved.
This is why we study elliptic curves with a point of order at least 4 separately.

\subsection{Universal Elliptic Curves with a Point of Order $n \geq 4$}

Let $E/K$ be an elliptic curve and let $P\in E(K)$ be a point which has order at least 4 in the MW group.
We can do a change of variable that takes the tangent at $P$ to the line $y = 0$.
Since the order of $P$ is not 2, the equation of the tangent has the form $y = \lambda x + \nu$, so the equation we obtain after this change of variables only contains ``Weierstrass monomials". 
Explicitly, if the tangent at $P$ is $y = \lambda x + \nu$,
then we set $y' = y-(\lambda x + \nu)$;
setting $y' = 0$ is then the same as requiring $y = \lambda x + \nu$.
Thus, in the new coordinates, the equation of $E$ has the form:
\begin{align}
 y^2 + a_1 xy + a_3 y = x^3 + a_2 x^2\, , \end{align}
since the polynomial on the RHS is the equation of $E$ restricted to the tangent at $(0,0)$,
and therefore has to vanish twice at $(0,0)$.
Finally, multiplying through by $\frac{a_2^6}{a_3^6}$ and replacing $x,y$ by $\frac{a_3^2x}{a_2^2}, \frac{a_3^3}{a_2^3} y$ gives a Weierstrass equation:
\begin{align} y^2 + \frac{a_1 a_2}{a_3} xy + \frac{a_2^3}{a_3^2} y = x^3 + \frac{a_2^3}{a_3^2} x^2 \end{align}
The main things to notice about the last equation are that only the coefficients $a_1,a_2,a_3$ are nonzero, and that $a_2 = a_3$.\\
Thus, for any elliptic curve $E$ with a point $P$ of order at least 4,
there exist unique $u,v \in K$ such that the pair $E,P$ is isomorphic to:
\begin{align}
 y^2 + (1-u)xy -v y = x^3 - vx^2 \, .
\end{align}
We note that we are using parameters $1-u, -v$ instead of just $u,v$ to simplify future computations.\\
Viewing this as an elliptic curve over $K(u,v)$,
we can compute multiples of $P = (0,0)$:
\[ P = (0,0) \, ,\qquad 2P = (v,uv)\, , \qquad 3P = (u, v-u)\, , \]
\[ -P = (0,v)\, , \qquad -2P = (v,0)\, , \qquad -3P = (u,u^2)\, , \]
\[ 4P =\left(\frac{ (v - u) v}{u^2}\, , \frac{v^2 (u^2 + u - v)}{u^3}\right)\, , \quad
5P = \left(\frac{uv(u^2 + u - v)}{(u - v)^2}\, , \frac{u^2 v (u^3 + u v - v^2)}{(u - v)^3}\right)\, , \]
\[ -4P =\left(\frac{ (v - u) v}{u^2}\, , \frac{(v-u)^2 v}{u^3}\right) \, ,\quad
-5P = \left(\frac{uv(u^2 + u - v)}{(u - v)^2}\, , \frac{v^2 (u^2 + u  - v)^2}{(u - v)^3}\right)\, . \] 
We can construct the universal elliptic curve with an $n$-torsion point for $n = 4,5,6$ by setting $P = - 3P, 3P = -2P, 3P = -3P$, respectively. 
\begin{align*}
4P = O& \Longleftrightarrow P = -3P \Longleftrightarrow u = 0 \, , \\
5P = O& \Longleftrightarrow 3P = -2P \Longleftrightarrow u = v \, ,\\
6P = O& \Longleftrightarrow 3P = -3P \Longleftrightarrow u^2+u = v \, ,\\
\end{align*}
This gives us the universal elliptic curves:
\begin{align}
y^2 + xy - ty = x^3 - tx^2 \, ,\qquad (0,0) \in E[4]\, , \\
 y^2 +(1-t)xy+ty = x^3 - tx^2 \, ,\qquad (0,0) \in E[5]\, ,\\
y^2 + (1-t)xy-(t^2+t)y = x^3-(t^2+t)x^2\, , \qquad (0,0) \in E[6]\, ,
\end{align}

Next, we compute the universal elliptic curves for $n = 7, 8$.
For these fibrations, the relation between $u,v$ will define a singular curve,
and we will compute the normalization of the curve explicitly to find the universal curve.\\
For $n = 7$, we set $5P = -2P$ to obtain:
\[7P = O\, , \quad \Longleftrightarrow u^3-uv+v^2 =0 \, . \]
The cubic is clearly nodal.
The normalization of the curve is:
\[ \spec \CC[t] \rightarrow C' = \spec \CC[u,v]/u^3-uv+v^2\, , \qquad t \mapsto (t^2-t,t^3-t^2) \, ,\]
and has a rational inverse given by $(u,v)\mapsto \frac{v}{u}$.
Thus, the universal elliptic curve is:
\[ y^2 + (1+t-t^2)xy+(t^2-t^3)y = x^3+(t^2-t^3)x^2\, . \] 
For $n =8$, we set $4P =-4P$ to obtain:
\[ v(u^2+u-v) =(v-u)^2 \,. \] 
The normalization of this curve is:
\[ t\mapsto \left(\frac{(2t-1)(t-1)}{t}, (2t-1)(t-1)\right)\, . \]
Note that the inverse of the normalization map is again $(u,v)\mapsto \frac{v}{u}$. 
\subsubsection{Extending $\phi$} 
Let $\pi : Y \rightarrow B$ be an elliptic fibration with an $n$-torsion section $P$, 
$4\leq n \leq 8$.
Let $S\rightarrow \PP^1$ be the N\'eron model of one of the following universal elliptic curves over $\CC(t)$:
\begin{align} y^2+(1+t-t^2)xy +(t^2-t^3)y &=x^3+(t^2-t^3) x^2 \, , \\
 y^2 +(t^2- (t-1)(2t-1))xy - t^3(t-1)(2t-1) y &= x^3 - t^2(t-1)(2t-1)x^2\, . \end{align}
Note that $(0,0)$ has order $n$ where $n= 7$ for the first equation and $n = 8$ for the second. 
Let $U = B - V(\Delta)$ and $Y_U =Y \times_B U$.
We have a commutative diagram:
\begin{equation}
\begin{tikzcd}
Y_U \arrow[r, "\Phi"] \arrow{d}{\pi}
&S \arrow{d}{p}\\
B_U \arrow[r, "\phi"] \arrow[u, bend right = 40] &\PP^1 \arrow[u, bend right = 40]\, ,
\end{tikzcd}
\end{equation}
with $B$ flat and non-constant.
We can take the closure of the graph to extend to a diagram:
\begin{equation}
\begin{tikzcd}
Y_0 \arrow[r, "\Phi"] \arrow{d}{\pi}
&S \arrow{d}{p}\\
B_0 \arrow[r, "\phi"] \arrow[u, bend right = 40] &\PP^1 \arrow[u, bend right = 40]\, .
\end{tikzcd}
\end{equation}
The map $B_0 \rightarrow \PP^1$ is obtained by composing the map to the singular modular curve (in $u,v$ coordinates):
\begin{align}
 b\mapsto \left( \frac{a_3-a_1a_2}{a_3}(b), \frac{-a_2^3}{a_3^2}(b) \right)\, , \end{align}
with the inverse of the normalization map.
The inverse of the normalization map is the same for both $n = 7,8$, 
and the composite map is:
\begin{align}
b\mapsto \left[\frac{a_3(a_3-a_1a_2)}{a_2^3}(b):1 \right]\, . \end{align}
\subsection{Torsion groups of type $\ZZ_2 \times \ZZ_{2m}$} 
In this section we derive the Weierstrass equation for the elliptic surface with torsion group $\ZZ_2 \times \ZZ_6$.\\
Any elliptic curve with torsion group $\ZZ_2\times \ZZ_6$ has a point of order 6, and therefore admits a Weierstrass equation of the form:
\begin{align}
 y^2 + (1-\psi(t))xy -(\psi(t)+\psi(t)^2)y = x^3-(\psi(t)+\psi(t)^2) x^2 \, , \end{align}
with $\psi(t)$ a function on the base.
To find the ``universal $\psi$", we use the following facts from algebra:
\begin{itemize}
\item{Let $G$ be a group. Then $G$ contains a subgroup isomorphic to $\ZZ_2 \times \ZZ_6$ if and only if it contains a subgroup isomorphic to $\ZZ_6$ and a subgroup isomorphic to $\ZZ_2 \times \ZZ_2$.}
\item{The 2-torsion points of an elliptic curve:
\begin{align}
 y^2 = x^3 + fx + g\, , \end{align}
are in bijection with roots of $x^3 + fx + g$.
The discriminant of the elliptic curve is the same (up to a unit) as the discriminant of the cubic.} 
\item{All roots of $x^3 + fx + g$ are defined over a field $K$ if and only if the cubic has at least one root in the field, and the discriminant of the cubic is a perfect square.}
\end{itemize}
Thus, to find the Weierstrass equation, we need to find $\psi$ so that the equation above is a perfect square.
The choice $\psi(t) = \frac{10-2t}{t^2-9}$ gives us a Weierstrass equation for the elliptic surface:
\begin{align} y^2 + \left( 1 - \frac{10-2t}{t^2-9} \right)xy + \frac{2(t-1)^2(t-5)}{(t^2-9)^2}y = x^3 +\frac{2(t-1)^2(t-5)}{(t^2-9)^2}x^2 \, ,
\end{align} 
with discriminant:
\begin{align}
\Delta = \frac{(t-9)^2(t-1)^6}{(t^2-9)^2} \, .
\end{align}
This model has the correct torsion group. 

We now perform a standard change of variable to obtain an integral model:
\begin{align}
y^2+(t^2+2t-19)xy+2 (-5 + t) (-1 + t)^2 (-9 + t^2)y = x^3+2 (-5 + t) (-1 + t)^2x^2 \, .
\end{align}
The degree of $a_1$ is 2 and the degree of $a_i$ is bounded above by $2i$ for $i = 2,3$;
thus the fundamental line bundle of the elliptic surface over $\PP^1$ has degree 2, as expected from Miranda's Table~\ref{tab:MWK3}. 
\subsection{Examples}
In this section,
we give Weierstrass models of singular Calabi-Yau 3-folds with Mordell-Weil torsion in  $\set{\ZZ_7, \ZZ_8, \ZZ_2 \times \ZZ_6}$.
To construct these 3-folds,
we used the method described in the introduction:
\begin{itemize}
    \item To obtain the Weierstrass model of the modular surface $S\rightarrow \PP^1$, we use the construction in the universal elliptic curve section.
    \item We chose the map $[s_0:s_1]\times[t_0:t_1]\mapsto[s_0t_0:s_1t_1]$ as $\phi$, although any other bi-degree (1,1) map would have given an equivalent 3-fold. 
    Note that we need a bi-degree (1,1) function because the $\Ll_{S/\PP^1} = \Oo(2)$ for both $7$ and 8 torsion. 
    Note that for our choice of $\phi$,
    the locus of indeterminacy is $[1:0] \times [0:1]$ and $[0:1] \times [1:0]$.
    
    \end{itemize} 
To obtain the equation for the 3-fold, we pulled back the sections of the elliptic surface by $\phi$ to obtain the coefficients shown below.

Note that both 3-folds have $(f,g,\Delta)$ vanishing to order exactly $(8,12,24)$ over $[1:0]\times [0:1]$ and $[0:1]\times[1:0]$, as predicted by our lemma.

\subsubsection{Explicit Equations}

Explicit equations for the 3-folds over $\PP^1 \times \PP^1$ are given below.
We use coordinates $[s_0:s_1]\times [t_0:t_1]$ for the base.

\begin{itemize}
\item{$\ZZ/7$: 
\begin{align}
 y^2 + (s_0^2t_0^2+s_0s_1t_0t_1-s_1^2t_1^2) xy + s_0^3s_1^2t_0^3t_1^2(s_0t_0-s_1t_1)y = x^3 +s_0t_0s_1^2t_1^2(s_0t_0-s_1t_1)x^2 \, ,\end{align} 
The coefficients $f,g$ and the discriminant $\Delta = 4f^3 + 27g^2$ are given below:
 \begin{equation}
   \begin{split}
   f&=  \frac{1}{48}-(s_0^2 t_0^2 - s_0 s_1 t_0 t_1 + s_1^2 t_1^2)\times \\& (s_0^6 t_0^6 + 
   5 s_0^5 s_1 t_0^5 t_1 - 10 s_0^4 s_1^2 t_0^4 t_1^2 - 
   15 s_0^3 s_1^3 t_0^3 t_1^3 + 30 s_0^2 s_1^4 t_0^2 t_1^4 - 
   11 s_0 s_1^5 t_0 t_1^5 + s_1^6 t_1^6) \, ,\\
   g& =  \frac{1}{864}  s_0^2 t_0^2 (s_0^{10} t_0^{10} + 6 s_0^9 s_1 t_0^9 t_1 - 15 s_0^8 s_1^2 t_0^8 t_1^2 - 
   46 s_0^7 s_1^3 t_0^7 t_1^3 + 174 s_0^6 s_1^4 t_0^6 t_1^4  - 
   222 s_0^5 s_1^5 t_0^5 t_1^5 \\ & + 273 s_0^4 s_1^6 t_0^4 t_1^6 - 
   486 s_0^3 s_1^7 t_0^3 t_1^7 + 570 s_0^2 s_1^8 t_0^2 t_1^8 - 
   354 s_0 s_1^9 t_0 t_1^9 + 117 s_1^{10} t_1^{10}) \, ,
    \end{split}
    \end{equation} 
\begin{align}
   \Delta = -\frac{ 1}{16}s_0^7 s_1^7 t_0^7 t_1^7 (s_0 t_0 - s_1 t_1)^7 (s_0^3 t_0^3 + 
   5 s_0^2 s_1 t_0^2 t_1 - 8 s_0 s_1^2 t_0 t_1^2 + s_1^3 t_1^3) \, .
   \end{align}    
} 
\item{$\ZZ/8$:
 \begin{equation}
   \begin{split}
   &y^2 + (s_0 s_1 t_0 t_1 + (s_0 t_0 - s_1 t_1) (2 s_1 t_1-s_0t_0)) xy + s_0 s_1^3 t_0 t_1^3 (s_0 t_0 - 2 s_1 t_1) (s_1 t_1 - s_0 t_0) y \\
    &= x^3 + s_1^2 t_1^2 (s_0 t_0 - 2 s_1 t_1) (s_1 t_1 - s_0 t_0)x^2  \, .
    \end{split}
    \end{equation} 
The coefficients $f,g$ and $\Delta$ are: 
 \begin{equation}
   \begin{split}
  f = &\frac{1}{48} (-s_0^8 t_0^8 + 16 s_0^7 s_1 t_0^7 t_1 - 96 s_0^6 s_1^2 t_0^6 t_1^2 + 
 288 s_0^5 s_1^3 t_0^5 t_1^3 - 480 s_0^4 s_1^4 t_0^4 t_1^4 + 
 448 s_0^3 s_1^5 t_0^3 t_1^5\\
 & - 224 s_0^2 s_1^6 t_0^2 t_1^6 + 
 64 s_0 s_1^7 t_0 t_1^7 - 16 s_1^8 t_1^8)\, , \\
   g =& \frac{1}{864}  (s_0^4 t_0^4 - 8 s_0^3 s_1 t_0^3 t_1 + 
   16 s_0^2 s_1^2 t_0^2 t_1^2 - 16 s_0 s_1^3 t_0 t_1^3 + 
   8 s_1^4 t_1^4) (s_0^8 t_0^8 - 16 s_0^7 s_1 t_0^7 t_1 + 
   96 s_0^6 s_1^2 t_0^6 t_1^2 \\
   & - 288 s_0^5 s_1^3 t_0^5 t_1^3 + 
   456 s_0^4 s_1^4 t_0^4 t_1^4 - 352 s_0^3 s_1^5 t_0^3 t_1^5 + 
   80 s_0^2 s_1^6 t_0^2 t_1^6 + 32 s_0 s_1^7 t_0 t_1^7 - 
   8 s_1^8 t_1^8)\, , \\ 
   \Delta = &\frac{-1}{16}s_0^2 s_1^8 t_0^2 t_1^8 (s_0 t_0 - 2 s_1 t_1)^4 (s_0 t_0 - 
   s_1 t_1)^8 (s_0^2 t_0^2 - 8 s_0 s_1 t_0 t_1 + 8 s_1^2 t_1^2) \, .
    \end{split}
    \end{equation}  } 
   \item{$\ZZ_2 \times \ZZ_6$:
   \begin{equation}
   \begin{split}
   &y^2+ (-19 s_0^2 t_0^2 + 2 s_0 s_1 t_0 t_1 + s_1^2 t_1^2) xy+ 2 s_0 t_0 (s_0 t_0 - s_1 t_1)^2 (5 s_0 t_0 - s_1 t_1) (9 s_0^2 t_0^2 - 
    s_1^2 t_1^2) y \\
    &= x^3 +2 s_0 t_0 (s_0 t_0 - s_1 t_1)^2 (5 s_0 t_0 - s_1 t_1) x^2  \, ,
    \end{split}
\end{equation}
 
    \begin{equation}
\begin{split}
 f &= \frac{1}{48} (-119761 s_0^8 t_0^8 + 36920 s_0^7 s_1 t_0^7 t_1 + 
    15700 s_0^6 s_1^2 t_0^6 t_1^2 + 11432 s_0^5 s_1^3 t_0^5 t_1^3 - 
    11958 s_0^4 s_1^4 t_0^4 t_1^4\\
    & + 1992 s_0^3 s_1^5 t_0^3 t_1^5 + 
    180 s_0^2 s_1^6 t_0^2 t_1^6 - 40 s_0 s_1^7 t_0 t_1^7 - s_1^8 t_1^8)\, , \\
g &= \frac{1}{864} (-41545241 s_0^{12} t_0^{12} + 19809780 s_0^{11} s_1 t_0^{11} t_1 + 
    4915350 s_0^{10} s_1^2 t_0^{10} t_1^2 + 7207028 s_0^9 s_1^3 t_0^9 t_1^3 \\
    &-9699039 s_0^8 s_1^4 t_0^8 t_1^4 + 2418984 s_0^7 s_1^5 t_0^7 t_1^5 + 
    172820 s_0^6 s_1^6 t_0^6 t_1^6 - 12216 s_0^5 s_1^7 t_0^5 t_1^7 \\
    &-56679 s_0^4 s_1^8 t_0^4 t_1^8 + 12388 s_0^3 s_1^9 t_0^3 t_1^9 - 
    330 s_0^2 s_1^{10} t_0^2 t_1^{10} - 60 s_0 s_1^{11} t_0 t_1^{11} - s_1^{12} t_1^{12}) \, , \\
\Delta &= 2 s_0^4 t_0^4 (s_0 t_0 - s_1 t_1)^6 (-5 s_0 t_0 + s_1 t_1)^4 (-9 s_0^2 t_0^2 + 
    s_1^2 t_1^2)^2 (2969 s_0^6 t_0^6 - 318 s_0^5 s_1 t_0^5 t_1  \\
   & -339 s_0^4 s_1^2 t_0^4 t_1^2 - 556 s_0^3 s_1^3 t_0^3 t_1^3 + 
    331 s_0^2 s_1^4 t_0^2 t_1^4 - 38 s_0 s_1^5 t_0 t_1^5 - s_1^6 t_1^6)
\end{split}
\end{equation} 
}
\end{itemize}

 \section{Index of Congruence Subgroups}
 \label{app:indices}
 
 In this section, we explain how to compute the indices $[\mathrm{SL}(2,\ZZ): \Gamma]$ where $\Gamma$ is one of the congruence subgroups defined earlier.
 Further details can be found in \cite{diamondshurman}.
 
 \subsection{$\Gamma(n)$}
 We start by computing the index of $[\mathrm{SL}(2,\ZZ) : \Gamma(n)]$ for $n > 1$.
The short exact sequence:
 \begin{align}
 0 \longrightarrow \Gamma(n) \longrightarrow \mathrm{SL}(2,\ZZ) \rightarrow \mathrm{SL}_2(\ZZ_n) \longrightarrow 0 \, ,\end{align}
 shows that $[\mathrm{SL}(2,\ZZ) : \Gamma(n)] = |\mathrm{SL}_2(\ZZ_n)|$.
 Furthermore, by the Chinese Remainder Theorem,
 we have isomorphisms:
 \begin{align} \mathrm{SL}_2(\ZZ_{nm}) \rightarrow \mathrm{SL}_2(\ZZ_n) \times \mathrm{SL}_2(\ZZ_m) \, , \end{align}
 whenever $n, m$ are relatively prime.
Thus, to compute $[\mathrm{SL}(2,\ZZ): \Gamma(n)]$ for general $n$,
it suffices to find a formula for $|\mathrm{SL}_2(\ZZ_n)|$ when $n$ is a prime power.\\
 We derive that formula in the following.
 Assume $n = p^k$ for some prime $p$ and some positive integer $k$.
 We will count the 4-tuples $(a,b,c,d) \in \ZZ_n^4$ with $ad- bc = 1$.

 \begin{itemize}
 \item{We count 4-tuples where $a$ is a unit first.
 There are $p^k - p^{k-1}$ units in $\ZZ_n$, so we have that many choices for $a$.
 Having chosen $a$, we can choose $b, c$ freely, and take $d = \frac{1+bc}{a}$ to guarantee the corresponding matrix has determinant 1.
 Altogether, there are:
 \begin{align}
 (p^k-p^{k-1})\cdot p^k \cdot p^k \cdot 1 = p^{3k}(1-\frac{1}{p}) \, ,
 \end{align}
 such 4-tuples.}
 \item{Next, we count 4-tuples where $a$ is not a unit.
 We can choose $a$ in one of $p^{k-1}$ ways.
 Since $a$ is not a unit, $b,c$ necessarily have to be units.
 We can choose $b$ freely in the set of units, and we can choose $d$ freely in $\ZZ_n$.
 We then solve for $c$ as $c = \frac{ad-1}{b}$.
This means the number of 4-tuples is:
 \begin{align}
 p^{k-1} \cdot (p^k - p^{k-1}) \cdot p^k \cdot 1 = p^{3k}(\frac{1}{p} - \frac{1}{p^2} ) \, ,
 \end{align}}
 \item{Altogether, this means we have:
 \begin{align} p^{3k}\left( \left( 1 - \frac{1}{p} \right) + \left( \frac{1}{p} - \frac{1}{p^2} \right) \right) = p^{3k}\left( 1 - \frac{1}{p^2} \right) \, , \end{align}
elements in $\mathrm{SL}_2(\ZZ_n)$.
}
 
 \end{itemize} 
 Finally, if $n = n_1 \cdots n_r$,
 with $n_i = p_i^{e_i}$, $p_i \neq p_j$ for $i \neq j$,
 then:
 \begin{align}
 [\mathrm{SL}(2,\ZZ) : \Gamma(n)] = \prod_i [\mathrm{SL}(2,\ZZ) : \Gamma(n_i)]  = n^3 \prod_i 1 - \frac{1}{p_i^2}\, .
 \end{align}
 \subsection{$\Gamma_1(n)$}
 
 Next we use the computation of $[\mathrm{SL}(2,\ZZ) : \Gamma(n)]$ to obtain formulas for the index of $\Gamma_1(n)$.
 Recall that elements of $\Gamma_1(n)$, after passing to the quotient, have the form $\ttm{1}{a}{0}{1}$ for some $a \in \ZZ_n$,
 and that:
 \begin{align}
 \ttm{1}{a}{0}{1} \ttm{1}{a'}{0}{1} = \ttm{1}{a+a'}{0}{1}  \, .
 \end{align}
 Thus, we have a short exact sequence:
 \begin{align}0 \rightarrow \Gamma(n) \rightarrow \Gamma_1(n) \rightarrow \ZZ_n \rightarrow 0 \, ,\end{align}
 where the map $\Gamma_1(n)\rightarrow \ZZ_n$ is the projection onto the top-right factor.
 This shows that $[\Gamma_1(n):\Gamma(n)] = n$, so we obtain:
 \begin{align}
 [\mathrm{SL}(2,\ZZ) : \Gamma_1(n)] = \frac{1}{n} [\mathrm{SL}(2,\ZZ) : \Gamma(n)]  = n^2 \prod_{p|n} 1 - \frac{1}{p^2} \, .
 \end{align} 
 We can use this formula to check one of the claims made in the main body - namely, that $24 | [\mathrm{SL}(2,\ZZ) : \Gamma_1(n)]$ for $n \geq 5$.
 By multiplicativity of the index, it suffices to prove the divisibility claim when $n$ is a prime power. 
 \begin{itemize} 
 \item{First, assume $n = p^k$ for $p \geq 5$ and $k \geq 1$.
 The formula above shows that:
 \begin{align}
 [\mathrm{SL}(2,\ZZ) : \Gamma_1(n)] = p^{3k} - p^{3k-2} = p^{3k-2}(p+1)(p-1) \, .
 \end{align}
 Since $3 \not| p$, $p^2 -1$ is necessarily divisible by 3.
 Furthermore, since $p$ is odd, both $p+1, p-1$ are even and exactly one of them is divisible by 4.
 Thus, $p^2 - 1$ is divisible by 3 and 8, so it is divisible by 24.}
 \item{Next, assume $n = 3^k$ for $k \geq 2$.
 Since $3$ is odd, $3^{2k} -1$ is divisible by 8 for the same reason as above.
 Furthermore, since $k > 1$, $3k-2 \geq 1$, so the index is divisible by 3,
 so again we see that the index is divisible by 24.}
 
 \item{Finally, assume $n = 2^k$.
 Then:
 \begin{align}
 [\mathrm{SL}(2,\ZZ): \Gamma_1(n)]  = 2^{3k} - 2^{3k-2} = 2^{3k-2} \cdot 3
 \end{align}
 In particular, the index is divisible by 24 precisely when $3k-2>3$,
 i.e. $k \geq 3$.}
 
 \end{itemize}
 The observations above prove the claim for almost all $n$.
 We can check the claim by hand for $n = 6, 12$,
 since they are not divisible by $8,9$ or any larger prime,
 to complete the argument.

 \subsection{$\Gamma_0(n)$}
 
 We compute $[\mathrm{SL}(2,\ZZ) : \Gamma_0(n)]$ using a similar argument.
 A general element of $\Gamma_0(n)$ reduces to a matrix of the form $\ttm{a}{b}{0}{d}$ in $\mathrm{SL}_2(\ZZ_n)$.
 The diagonal entries are necessarily units after we reduce mod $n$.
 Now, observe that:
 \begin{align} \ttm{a}{b}{0}{d} \ttm{a'}{b'}{0}{d'} = \ttm{aa'}{ab'+bd'}{0}{dd'} \, . \end{align}
 Thus, we can write down a group homomorphism $\Gamma_0(n)\rightarrow \ZZ_n^\times$ whose kernel is exactly $\Gamma_1(n)$,
 so $[\Gamma_0(n) : \Gamma_1(n)] = \phi(n)$.
 Using the identity $\phi(n) = n \prod_{p|n} 1 - \frac{1}{p}$, 
 we obtain:
 \begin{align}
 [\mathrm{SL}(2,\ZZ): \Gamma_0(n)] = \frac{1}{\phi(n)} [\mathrm{SL}(2,\ZZ): \Gamma_1(n)] = n \prod_{p|n} 1 +\frac{1}{p} \, .
 \end{align}
 \subsection{Non-classical congruence groups}
 
 For completeness, we include a formula for the index of the group that parametrizes $(E, (P,Q))$ for $E$ an elliptic curve, $P$ a point of order $nd$ and $Q$ a point of order $d$ for integers $n,d$.
 
 The idea is to take the fiber product of $X_1(nd)$ and $X(d)$.
 At the level of groups, this corresponds to taking the intersection of the groups. 
To compute the index of the intersection, we use the identity:
\begin{align}
[\mathrm{SL}(2,\ZZ):\Gamma_1(nd) \cap \Gamma(d)] = [\mathrm{SL}(2,\ZZ) : \Gamma_1(nd)][\Gamma_1(nd):\Gamma_1(nd) \cap \Gamma(d) ] \, .
\end{align}
 An element of $\Gamma_1(nd)$ has the form $\ttm{1}{a}{0}{1}$ mod $nd$.
 In order for it to be an element of $\Gamma(d)$, $a$ has to be a multiple of $d$.
 Thus, the index of the intersection in $\Gamma_1(nd)$ is $n$
 so:
\begin{align}
 [\mathrm{SL}(2,\ZZ):\Gamma_1(nd) \cap \Gamma(d)] = n^3d^2 \prod_{p | nd} 1- \frac{1}{p^2} \, .
\end{align} 
Thus, the index of the group associated to the Mordell-Weil torsion $\ZZ_2 \times \ZZ_4$ has index 24, $\ZZ_2 \times \ZZ_6$ has index 48, $\ZZ_3 \times \ZZ_6$ has index 72 and $\ZZ_2 \times \ZZ_8$ has index 96.
Applying the formula for the degree of the fundamental line bundle for $\Gamma_1(n)$, this predicts the corresponding universal surface has a fundamental line bundle of degree 1,2,3,4, respectively.
These numbers match up exactly with the table in \cite{MP}.
 
 We have not discussed torsion groups of type $\ZZ_m \times \ZZ_n$ where $n \neq m$ and neither divides the other, but the formulas above cover all of the 19 subgroups with a quotient of genus 0.


\end{document}